\documentclass{article}
\newif\ifclean
\cleantrue  
\cleanfalse 
\usepackage{arxiv}
\usepackage{float}
\usepackage[square,sort,comma,numbers]{natbib}
\usepackage{graphicx}	
\usepackage{array}
\usepackage{adjustbox}
\usepackage{hyperref}
\usepackage{url}
\usepackage{amsmath,amssymb,amsthm}
\usepackage{mleftright,mathtools}
\usepackage{bm}
\usepackage[usenames,dvipsnames]{xcolor}

\usepackage[normalem]{ulem}
\newcommand{\COMMENT}[1]{\textcolor{cyan}{{[ \sc{#1} ]}}} 

\newcommand{\QUESTION}[1]{\textcolor{ForestGreen}{{#1}}}
\newcommand{\red}[1]{\textcolor{red}{{#1}}}

\newcommand{\ts}[1]{\boldsymbol{#1}}			

\newcommand{\tr}{\mathrm{tr}}

\newcommand{\eps}{\ts{\epsilon}}
\newlength{\figwidth}
\newlength{\figwidthtwo}
\newlength{\figwidththree}
\newcommand{\cref}[1]{Ref.\,\cite{#1}}

\newcommand{\Ic}{\mathcal{I}}

\newcommand{\Dc}{\mathcal{D}}

\newcommand{\varepsilonb}{\boldsymbol{\varepsilon}}

\newcommand{\bb}{\mathbf{b}}
\newcommand{\eb}{\mathbf{e}}

\newcommand{\nb}{\mathbf{n}}

\newcommand{\db}{\mathbf{d}}

\newcommand{\tb}{\mathbf{t}}

\newcommand{\xb}{\mathbf{x}}

\newcommand{\pb}{\mathbf{p}}
\newcommand{\ub}{\mathbf{u}}
\newcommand{\Ab}{\mathbf{A}}
\newcommand{\Bb}{\mathbf{B}}
\newcommand{\Cb}{\mathbf{C}}

\newcommand{\Gb}{\mathbf{G}}

\newcommand{\Fb}{\mathbf{F}}

\newcommand{\Eb}{\mathbf{E}}

\newcommand{\Nb}{\mathbf{N}}
\newcommand{\Pb}{\mathbf{P}}
\newcommand{\Sb}{\mathbf{S}}
\newcommand{\Ib}{\mathbf{I}}
\newcommand{\Rb}{\mathbf{R}}
\newcommand{\Qb}{\mathbf{Q}}

\newcommand{\Xb}{\mathbf{X}}

\newcommand{\bX}{\mathbf{X}}

\newcommand{\Grad}{\nabla_{\!\bX}}   

\usepackage{booktabs}
\usepackage{tabularx}
\usepackage[letterpaper]{geometry}
\usepackage{subcaption}
\graphicspath{{./Figures/}}

\usepackage{comment}
\newcommand{\caution}{\red{\bf Draft: \today. Do not distribute.}}
\ifclean
\renewcommand{\COMMENT}[1]{{}}
\renewcommand{\QUESTION}[1]{{}}
\else
\fi

\usepackage[utf8]{inputenc}
\usepackage[english]{babel}
\usepackage{natbib}
\title{\bf Multiscale topology optimization of compressible and nearly incompressible anisotropic hyperelastic structures using physics-augmented neural networks}
\renewcommand{\shorttitle}{\textbf{Multiscale topology optimization using physics-augmented neural networks}}

\author{
\textbf{Asghar A. Jadoon}$^{1,*}$ \qquad
\textbf{Aryan Tyagi}$^{1,*}$ \qquad
\textbf{L. River Spencer}$^{1}$ \qquad
\textbf{Reese E. Jones}$^{2}$ \\[0.4em]
\textbf{Manuel K. Rausch}$^{1,3,4}$ \qquad
\textbf{Ryan Alberdi}$^{5}$ \qquad
\textbf{D. Thomas Seidl}$^{5}$ \qquad
\textbf{Jan N. Fuhg}$^{1,4,\dagger}$ \\[1em]
\begin{minipage}{0.9\textwidth}
\raggedright
\small
$^{1}$Department of Aerospace Engineering \& Engineering Mechanics, The University of Texas at Austin, Austin, TX 78712\\
$^{2}$Sandia National Laboratories, Livermore, CA 94551\\
$^{3}$Department of Biomedical Engineering, The University of Texas at Austin, Austin, TX 78712\\
$^{4}$The Oden Institute of Computational Science and Engineering, The University of Texas at Austin, Austin, TX 78712\\
$^{5}$Sandia National Laboratories, Albuquerque, NM 87185\\[0.5em]
$^{*}$These authors contributed equally to this work.\\
$^{\dagger}$Correspondence: \texttt{jan.fuhg@utexas.edu}
\end{minipage}
}

\date{}
\begin{document}

\maketitle

\section*{Abstract}
Multiscale topology optimization (TO) of hyperelastic materials remains computationally prohibitive due to the repeated solution of microscale boundary value problems. In this work, we present a concurrent multiscale topology optimization framework that overcomes this limitation by leveraging physics-augmented neural networks (PANNs) as surrogate constitutive models. The proposed approach enables the simultaneous optimization of macroscale material distribution and microscale descriptors, within a unified nonlinear finite strain setting. The surrogate models are constructed using input-specific neural networks (ISNNs) that enforce key physical principles directly within the architecture, including convexity and material symmetry through invariant-based representations and structural tensors. This ensures thermodynamic consistency and numerical stability while accurately representing homogenized anisotropic hyperelastic responses. The trained PANNs replace the microscale boundary value problem and provide efficient evaluations of stresses and consistent tangent moduli using analytical first and second derivatives of the neural network, enabling tractable large-scale multiscale optimization. The framework is demonstrated on representative microstructures exhibiting transversely isotropic, cubic anisotropic, and nearly incompressible isotropic behavior. The results show that the proposed method captures complex multiscale interactions and enables physically meaningful spatial tailoring of material properties, while significantly reducing computational cost compared to classical FE$^2$ approaches. These findings establish PANNs as a powerful tool for high-fidelity multiscale topology optimization of nonlinear anisotropic materials.

\section{Introduction}
\label{sec:introduction}
Advances in computational modeling have increased the use of topology optimization (TO) in a variety of fields such as solid mechanics \cite{TO_solids_1}, fluid dynamics \cite{TO_fluids_1}, and electromagnetics \cite{TO_elec_1}. In many of these applications, structures operate beyond the small-strain linear regime. Thus, nonlinear material behavior,  including regimes characterized by near incompressibility,  is often unavoidable under large deformations and/or inelastic effects, and single-scale nonlinear TO has therefore received significant attention \cite{Han2025elastoplasticfinite, kim2024elastoplasticfinite, zhang2022elastoplasticfinite, alberdi2017elastoplastic} with several works focused on incompressible materials \cite{khandelwal2018incomp, li2016incomp, bruggi2009incomp}. 
However, in functionally graded materials, metamaterials, or bio-inspired structures, the macroscopic performance is intrinsically tied to the material's complex microstructural features. Therefore, the exclusive design of the macroscale topology might be insufficient to achieve multifunctional performance, such as tunable anisotropy \cite{zheng2021tunable}, enhanced energy absorption \cite{martinez2025energy}, or locally tailored stiffness \cite{islam2024fiber}.
Multiscale TO addresses this need by allowing simultaneous control over both macroscale and microscale geometry \cite{sigmund1994inverse, sigmund1995inverse, wu2021review}. This capability is particularly relevant in emerging fields like architected materials \cite{architected, magneto_active}, additive manufacturing \cite{AM1, AM2}, and lattice-based designs \cite{lattice,lattice_}.

Extending single-scale nonlinear TO to multiple scales has been difficult in practice, even if a separation of scales \cite{fish2010multiscale} can be assumed. Classical multiscale methods typically involve multilevel Finite Element calculations (FE$^2$) \cite{FE2, FE2_2} where at every macroscale integration point a full Finite Element (FE) problem must be solved on a representative volume element (RVE) or a statistical volume element (SVE) to obtain stresses, tangent moduli, and sensitivities. As a result, a single iteration of the optimization algorithm requires hundreds of thousands of FE solutions (the order of discretization of the nominal geometry), making the procedure prohibitively expensive for nonlinear, and especially nearly incompressible, materials or large-scale designs. In particular, in applications involving soft materials such as rubbers \cite{rubber} and biological tissues \cite{bio_tissue,soft_tissue_rubber}, the nonlinearity is often accompanied by incompressibility, which introduces additional volumetric deformation that must be satisfied in addition to thermodynamic consistency and material symmetries. Furthermore, the issue of volumetric locking requires the use of mixed pressure-displacement formulations to obtain accurate and stable solutions \cite{sussman1987incompformulation}. Some other studies have considered topology optimization of incompressible or nearly incompressible materials under large deformations, primarily within single-scale formulations \cite{khandelwal2018incomp, li2016incomp, bruggi2009incomp}. Thus, much of the multiscale TO research in the past has been restricted to linearly elastic materials \cite{linear_1, linear_2, linear_3, sigmund2021ortho}. This is due to the fact that it is easier to build surrogates for linear homogenized responses and that modeling nonlinear materials introduces an additional nonlinear solver (Newton-Raphson) loop at each material point. Moreover, obtaining adjoints for nonlinear and time- and history-dependent constitutive models is non-trivial \cite{alberdi2018sensitivity}. 
Although applied to material variability uncertainty quantification rather than TO, the microstructural latent space approach of Ref.~\cite{jones2024multiscale} is adaptable to multiscale structural optimization.

Nevertheless, numerous attempts have been made to solve multiscale nonlinear TO problems by proposing workarounds to reduce the computational burden of FE$^2$. Ref. \cite{fritzen2016elastoviscoplastic} proposed a framework for small-strain elastoplastic structures using an intrinsic reduced-order model to speed up the evaluation of the microstructure. Ref. \cite{kato2018decoupling} focuses on reducing the computational cost of large-deformation multiscale TO by decoupling the micro and macroscale during homogenization. They solve the macroscale FE problem with an assumed phenomenological constitutive law whose parameters are identified from ``numerical material tests" of microstructural RVEs.  Another recent area to alleviate the computational bottlenecks of repeated homogenization has explored data-driven surrogate models, including neural networks (NNs) \cite{deeponet, path_dependent} and Gaussian processes \cite{frankel2020tensor,fuhg2022physics}. These models learn the effective constitutive response directly from precomputed microscale simulations. Once trained, the surrogates act as the constitutive model, enabling rapid stress and tangent evaluations. 

However, a key limitation of conventional black-box ML constitutive models is that they do not incorporate established physical principles and mechanistic assumptions, an issue that becomes particularly critical in data-scarce regimes \cite{fuhg2024review}. 
Ref. \cite{deng2025data} proposed a data-driven TO framework for the design of nonlinear multiscale soft functionally graded materials under large deformations. Their approach uses Gaussian process regression to learn the microstructure-property map from precomputed microscale data while NNs parameterize the spatially varying design fields during optimization. However, specifically in nonlinear materials, failure to satisfy concepts such as thermodynamic consistency, frame invariance, or material symmetries can lead to nonphysical predictions and numerical difficulties when embedded in an optimization loop \cite{linden2023neural}. 
Motivated by these concerns, recent efforts have incorporated mechanistic principles and assumptions directly into the NN architectures, resulting in what have been referred to as physics-augmented neural networks (PANNs) \cite{kalina2023FEANN,kalina2024magneto,jadoon2025inverse,holthusen2026complement}. These PANNs rely on implicit enforcement of functional constraints guaranteed by input convex neural networks \cite{ICNN} (ICNNs) or input-specific neural networks \cite{ISNN} (ISNNs) that enforce certain functional constraints such as convexity or monotonicity. In this context, Ref.~\cite{bessa2025} were the first to integrate an ICNN-based homogenization constitutive model for the simultaneous optimization of material density and particle volume fraction in a multiscale TO setting. Notably, the study has been restricted to isotropic and compressible materials, leaving anisotropic and incompressible multiscale TO largely unexplored. 
On the other hand, on the macroscale, several works have investigated concurrent optimization of topology and fiber orientation in composite structures, including continuous-angle formulations, stress-constrained settings, and extensions to finite-deformation hyperelasticity \cite{islam2024fiber, silva2020fiber, silva2023fiber}. They commonly rely on a transversely isotropic continuum constitutive law. Thereby, they do have an explicit dependence on the microstructure. 
Based on these ideas, we propose a concurrent multiscale TO framework based on PANNs for anisotropic finite strain, as well as isotropic nearly incompressible, finite strain materials.

The proposed framework consists of two key components:
\begin{enumerate}
    \item A physics-augmented neural constitutive model using ISNNs, which is trained offline to predict the microstructure-dependent homogenized \textbf{anisotropic} constitutive response, including a consistent treatment of isotropic \textbf{nearly incompressible} behavior.
    \item  A concurrent multiscale topology optimization formulation that incorporates microstructural descriptors as additional design variables within the nonlinear optimization setting.
\end{enumerate}

\begin{figure}[htbp]
    \centering
    \includegraphics[width=1.0\linewidth]{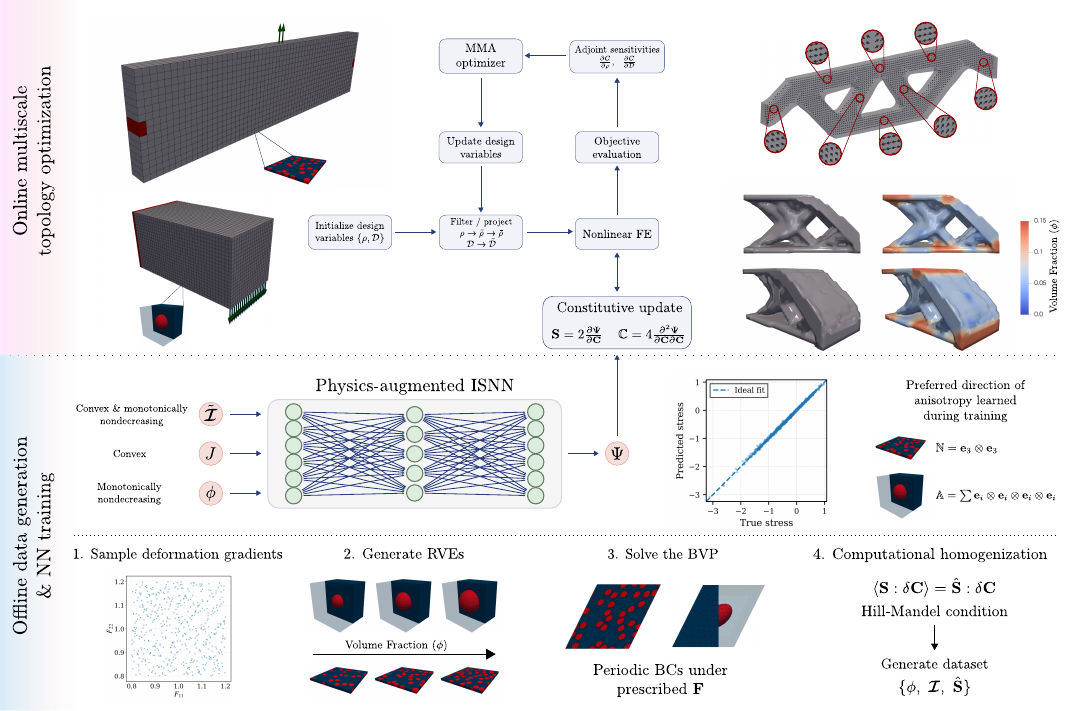}
    \caption{Overview of the proposed multiscale topology optimization framework. 
    The bottom panel illustrates the offline data generation and training of the physics-augmented neural network (PANN) surrogate based on computational homogenization. The top panel shows the online topology optimization loop, where the trained surrogate replaces the microscale boundary value problem and provides constitutive responses within a nonlinear finite element analysis.}
    \label{fig:overview}
\end{figure}

To our knowledge, no previous multiscale TO framework implements concurrent optimization of material density, degree of anisotropy, and preferred direction of anisotropy for finite strain regimes, while also consistently treating nearly incompressible material behavior across scales.

To provide an overview of the proposed framework, a schematic representation of the multiscale topology optimization approach is shown in Fig.~\ref{fig:overview}. The figure illustrates both the offline stage, where physics-augmented neural networks are trained using computational homogenization data, and the online stage, where the trained surrogate is embedded within a nonlinear finite element-based topology optimization loop. This schematic is intended to guide the reader through the overall methodology, whereas the individual components, including the constitutive modeling framework, surrogate construction, and optimization formulation, are described in detail in the subsequent sections. 

The remainder of this paper is organized as follows. Section \ref{Sec:Theory} discusses the theory and mathematical modeling of hyperelastic materials, and Section \ref{Sec:Model} introduces the PANN model architecture and training procedure. Section \ref{Sec:TO_problem} presents the full formulation of the multiscale topology optimization problem. Section \ref{Sec:NumericalExamples} provides numerical examples. Finally, the conclusion and discussion are given in Section \ref{Sec:discussion}.

\section{Constitutive framework}\label{Sec:Theory}

Let $\Omega_0 \subset \mathbb{R}^3$ denote the reference configuration occupied by a deformable body with material coordinates $\Xb \in \mathbb{R}^3$, and let the deformation mapping be $\xb(\Xb) = \Xb + \ub(\Xb)$, where $\ub$ is the displacement field. The mechanical response of hyperelastic materials is derived from the Helmholtz free energy $\Psi(\Fb)$ where $\Fb$ is the deformation gradient defined as $\Fb = \Ib + \Grad \ub$, where $\Grad$ denotes the gradient with respect to material coordinates in the reference configuration. We can also postulate a parametrized version of the free energy, i.e., $\Psi(\Fb, \bm\Dc)$ where $\bm\Dc$ denotes additional material or microstructural parameters. Then, through thermodynamic considerations, we can derive the first Piola-Kirchhoff stress as:

\begin{equation} \label{eq:P_def}
\Pb = \frac{\partial \Psi(\Fb, \bm\Dc)}{\partial \Fb} \ .
\end{equation}

The constitutive response is therefore fully determined by the choice of the free energy function $\Psi$. Alternatively, the free energy can be formulated in terms of the right Cauchy-Green deformation tensor $\Cb =\Fb^T\Fb$ such that the strain energy becomes:

\begin{equation}\label{eq:psi_C}
    \Psi = \Psi(\Cb, \bm\Dc) \, .
\end{equation}

This representation is particularly convenient, as it automatically satisfies frame indifference, which requires:

\begin{equation}\label{eq:objectivity}
    \Psi(\Qb \Fb, \bm\Dc) = \Psi(\Fb, \bm\Dc) \quad \forall \, \Qb \in \text{Orth}^{+} \ .
\end{equation}

Since $\Cb(\Qb\Fb)=\Cb(\Fb)$, objectivity is automatically satisfied. The work-conjugate stress measure associated with $\Cb$ is the second Piola-Kirchhoff stress, given by:

\begin{equation}\label{eq:S_def}
    \Sb = 2 \frac{\partial \Psi(\Cb, \bm\Dc)}{\partial \Cb} \, .
\end{equation}

This relation follows from the energetic conjugacy between the Green-Lagrange strain tensor $\Eb = \tfrac{1}{2}(\Cb - \Ib)$ and the second Piola-Kirchhoff stress $\Sb$. The two stress measures are related through $\Pb = \Fb \Sb$. In the following, we specialize this general framework to two formulations corresponding to compressible anisotropic and nearly incompressible isotropic materials.

\subsection{Compressible anisotropic hyperelasticity}

We consider a compressible anisotropic hyperelastic material within the general framework introduced above. In addition to objectivity, the free energy must satisfy material symmetry, which requires invariance under a symmetry group $\mathcal{G}$ acting on the reference configuration \cite{ogden1997non}. This is expressed as:

\begin{equation}\label{eq:material_symmetry}
    \Psi(\Fb \Gb^T, \bm\Dc) = \Psi(\Fb, \bm\Dc) \ , \qquad\Psi(\Gb \Cb \Gb^T, \bm\Dc) = \Psi(\Cb, \bm\Dc) 
    \quad \forall \, \Gb \in \mathcal{G} \subseteq \text{Orth}.
\end{equation}

Rather than constructing distinct functional forms of $\Psi$ for each symmetry group, it is customary to introduce structural tensors that encode the material symmetry \cite{zhang1990structural,svendsen1994representation}. Specifically, the free energy can be written as:
\begin{equation}\label{eq:psi_structural}
    \Psi = \Psi(\Cb, \bm{\mathcal{A}}, \bm\Dc),
\end{equation}
where $\bm{\mathcal{A}} = \{\Ab_i\}$ is a set of structural tensors. By definition, the structural tensors characterize the symmetry group $\mathcal{G}$ through the symmmetry condition:
\begin{equation}
    \Gb \Ab_i \Gb^T = \Ab_i 
    \quad \forall \, \Gb \in \mathcal{G} \ .
\end{equation}
The structural tensors are defined in the reference configuration, are independent of the deformation, and provide an equivalent representation of the material symmetry constraints to an inherent representation. By virtue of the \emph{isotropization} theorem \cite{zhang1990structural} and representation theorems for isotropic tensor functions \cite{gurtin1982introduction}, the free energy can be equivalently expressed as a function of a set of scalar invariants of $\Cb$ and $\bm{\mathcal{A}}$:
\begin{equation}\label{eq:invariant_psi}
    \Psi = \Psi(\bm{\mathcal{I}}, \bm\Dc),
\end{equation}
where:
\begin{equation}
    \bm{\mathcal{I}} = \bm{\mathcal{I}}_{\text{iso}}(\Cb) \cup \bm{\mathcal{I}}_{\text{aniso}}(\Cb, \bm{\mathcal{A}}).
\end{equation}
The isotropic invariants are given by:
\begin{equation}
    \bm{\mathcal{I}}_{\text{iso}} = 
    \left\{ 
    \Ic_1 = \tr(\Cb), \quad 
    \Ic_2 = \tr(\text{Cof} \,\Cb), \quad 
    \Ic_3 = J = \text{det}\Fb
    \right\}.
\end{equation}

\paragraph{Cubic anisotropy}

A fourth-order structural tensor is required to represent cubic anisotropy (see Ref.~\cite{kambouchev2007polyconvex} and references therein). When the axes of symmetry are aligned with the canonical basis, it can be defined as \cite{kambouchev2007polyconvex}:
\begin{equation}
    \tilde{\mathbb{A}} = \sum_{i=1}^3 \mathbf{e}_i \otimes \mathbf{e}_i \otimes \mathbf{e}_i \otimes \mathbf{e}_i.
\end{equation}

For a general orientation, the structural tensor is obtained via rotation:
\begin{equation}\label{eq:cubic_ST}
    {\mathbb{A}}= \sum_{i=1}^3 \Rb \mathbf{e}_i \otimes  \Rb\mathbf{e}_i \otimes  \Rb\mathbf{e}_i \otimes  \Rb\mathbf{e}_i
\end{equation}
Clearly, ${\mathbb{A}} = \tilde{\mathbb{A}}$ for $\Rb=\Ib$. The associated anisotropic invariants are defined as \cite{kambouchev2007polyconvex}:
\begin{equation}
    \bm{\mathcal{I}}_{\text{cub}} =
    \left\{
    \mathcal{I}_4 = \Cb : \mathbb{A} : \Cb, \quad
    \mathcal{I}_5 = \Cb : \mathbb{A} : \Cb^2, \quad
    \mathcal{I}_6 = \Cb^2 : \mathbb{A} : \Cb^2
    \right\}.
\end{equation}

\paragraph{Transverse isotropy}

While a fourth-order structural tensor can be used to model the behavior of transversely isotropic materials, similar to the case of cubic anisotropy, a simpler but sufficient way to represent transverse isotropy is through a second-order structural tensor, which we denote with $\mathbb{N}$ and can be obtained via $\mathbb{N} = \mathbf{n} \ \otimes \ \mathbf{n}$ where $\mathbf{n}$ is a unit vector representing the preferred direction of anisotropy.

For arbitrary orientation, $\mathbf{n}$ is obtained through rotation of the axis of symmetry $\mathbf{e}$ in the reference configuration:
\begin{equation}\label{eq:TI_ST}
    \mathbb{N}(\Rb) = \Rb \mathbf{e} \otimes \Rb \mathbf{e}.
\end{equation}

The associated invariants are given by \cite{schroder2008anisotropic}:
\begin{equation}
    \bm{\mathcal{I}}_{\text{TI}} =
    \left\{
    \mathcal{I}_4 = \tr(\Cb \mathbb{N}), \quad
    \mathcal{I}_5 = \tr((\text{Cof}\Cb)\mathbb{N})
    \right\}.
\end{equation}

From the invariant-based representation in Eq. \eqref{eq:invariant_psi}, the second Piola-Kirchhoff stress follows as:
\begin{equation}\label{eq:S_invariant}
\Sb 
= 2 \frac{\partial \Psi}{\partial \Cb}
= 2 \sum_i \partial_{\Ic_i} \Psi \; \partial_{\Cb} \Ic_i
= \sum_i c_i(\bm{\mathcal{I}}, \bm\Dc)\, \Bb_i,
\end{equation}
where $\Bb_i = \partial_{\Cb} \Ic_i$ are known tensor bases and $c_i = \partial_{\Ic_i}\Psi$ are scalar coefficient functions. Since the functional form of the invariants is known \textit{a priori}, the corresponding basis tensors can be computed analytically. As these tensors are symmetric, the resulting stress tensor $\Sb$ is also symmetric.

The governing equilibrium boundary value problem in the reference configuration is: find $\ub : \Omega_0 \rightarrow \mathbb{R}^3$ such that:
\begin{equation}\label{eq:strong_form_comp}
    \mathrm{Div}(\Fb \Sb) + \bb_0 = \mathbf{0} \quad \text{in } \Omega_0,
\end{equation}
where $\mathrm{Div}(\cdot)$ denotes the divergence operator with respect to the reference configuration, $\Fb \Sb = \Pb$ is the first Piola-Kirchhoff stress tensor, and $\bb_0$ is the body force per unit reference volume. The boundary $\partial \Omega_0$ is decomposed into a displacement boundary $\Gamma_{0,u}$ and a traction boundary $\Gamma_{0,t}$ with $\Gamma_{0,u} \cap \Gamma_{0,t} = \emptyset$ and $\Gamma_{0,u} \cup \Gamma_{0,t} = \partial \Omega_0$. The boundary conditions are given by:
\begin{equation}
    \ub = \bar{\ub} \quad \text{on } \Gamma_{0,u}, 
    \qquad 
    \Fb \Sb \Nb = \bar{\tb}_0 \quad \text{on } \Gamma_{0,t},
\end{equation}
where $\Nb$ is the outward unit normal to $\partial \Omega_0$ in the reference configuration and $\bar{\tb}_0$ is the prescribed traction per unit reference area. 

The solution follows from the principle of minimum potential energy. The weak form of the problem is: find $\ub$ such that for all admissible variations $\delta \ub$,
\begin{equation}\label{eq:weak_form_comp}
\int_{\Omega_0} \Sb : \delta \Eb \, dV
=
\int_{\Omega_0} \bb_0 \cdot \delta \ub \, dV
+
\int_{\Gamma_{0,t}} \bar{\tb}_0 \cdot \delta \ub \, dA,
\end{equation}
where $\Eb = \tfrac{1}{2}(\Cb - \Ib)$. The left-hand side represents the internal virtual work arising from the constitutive response of the material, while the right-hand side corresponds to the external virtual work due to body forces and prescribed tractions. Implementation-specific modifications to the kinematics and constitutive response for topology optimization are introduced in Section \ref{Sec:TO_problem}.

\subsection{Nearly incompressible isotropic hyperelasticity}

While the above formulation models volumetric response directly through the free energy, nearly incompressible materials require a different treatment of volumetric deformation, which is achieved through a mixed formulation. In this section, we restrict ourselves to isotropic materials and therefore omit any dependence on structural tensors. Furthermore, we consider a non-parametrized strain energy function.

To separate volumetric and distortional effects, we introduce the isochoric right Cauchy-Green tensor:

\begin{equation}\label{eq:C_bar}
    \bar{\Cb} = J^{-2/3}\Cb,
\end{equation}

with associated invariants:

\begin{equation}
    \bar I_1 = \tr(\bar{\Cb}), 
    \qquad
    \bar I_2 = \tfrac{1}{2}\big((\tr \bar{\Cb})^2 - \tr(\bar{\Cb}^2)\big).
\end{equation}

The free energy is additively decomposed into isochoric and volumetric contributions:

\begin{equation}
    \Psi(\Cb) = \Psi_{\mathrm{iso}}(\bar{\Cb}) + \Psi_{\mathrm{vol}}(J).
\end{equation}

The pressure associated with volumetric deformation is defined as:
\begin{equation}
    p = -\frac{\partial \Psi_{\mathrm{vol}}(J)}{\partial J}.
\end{equation}

To avoid volumetric locking in the nearly incompressible regime, we introduce an independent pressure field $\tilde p$ and define the free energy density as:
\begin{equation}\label{eq:psi_mixed}
    \Psi(\Cb,\tilde p)
    =
    \Psi_{\mathrm{iso}}(\bar{\Cb})
    -
    \tilde p (J-1)
    -
    \frac{1}{2\kappa}\tilde p^2,
\end{equation}

where the isochoric energy $\Psi_{\mathrm{iso}}(\bar{\Cb})$ is expressed in terms of invariants of $\bar{\Cb}$ and $\kappa$ denotes the bulk modulus.

The second Piola-Kirchhoff stress follows from the free energy as:
\begin{equation}\label{eq:S_mixed}
\Sb(\Cb,\tilde p)
=
2 \frac{\partial \Psi}{\partial \Cb}
=
\Sb_{\mathrm{iso}} - \tilde p\,J\,\Cb^{-1}.
\end{equation}
where the isochoric contribution is given by:
\begin{equation}\label{eq:S_iso}
    \Sb_{\mathrm{iso}} = 2 \frac{\partial \Psi_{\mathrm{iso}}}{\partial \Cb}.
\end{equation}

The governing boundary value problem in the reference configuration is: find $(\ub,\tilde p)$ such that
\begin{equation}\label{eq:strong_u}
    \mathrm{Div}(\Fb \Sb) + \bb_0 = \mathbf{0}
    \quad \text{in } \Omega_0,
\end{equation}
\begin{equation}\label{eq:strong_p}
    -(J-1) - \frac{\tilde p}{\kappa} = 0
    \quad \text{in } \Omega_0,
\end{equation}
where $\bb_0$ is the body force per unit reference volume. From Eq. \eqref{eq:strong_p}, the pressure field satisfies:

\begin{equation}
\tilde p = -\kappa (J - 1) \ ,
\end{equation}

which recovers the standard volumetric constitutive relation in the limit of the mixed formulation. The boundary conditions are given by:

\begin{equation}
    \ub = \bar{\ub} \quad \text{on } \Gamma_{0,u}, 
    \qquad 
    \Fb \Sb \Nb = \bar{\tb}_0 \quad \text{on } \Gamma_{0,t}.
\end{equation}

The solution follows from the stationarity of the total potential energy. The weak form of the problem is: find $(\ub,\tilde p)$ such that for all admissible variations $(\delta \ub, \delta \tilde p)$,
\begin{subequations}\label{eq:weak_mixed}
\begin{align}
\int_{\Omega_0} \Sb : \delta \Eb \, dV
&=
\int_{\Omega_0} \bb_0 \cdot \delta \ub \, dV
+
\int_{\Gamma_{0,t}} \bar{\tb}_0 \cdot \delta \ub \, dA, \\
\int_{\Omega_0} \left[ -(J-1) - \frac{\tilde p}{\kappa} \right] \delta \tilde p \, dV
&= 0,
\end{align}
\end{subequations}
where $\Eb = \tfrac{1}{2}(\Cb - \Ib)$. The first equation represents the balance of internal and external virtual work, while the second enforces the volumetric constraint associated with near-incompressibility. Implementation-specific modifications to the kinematics and constitutive response for topology optimization are introduced in Section \ref{Sec:TO_problem}.

\section{PANNs as surrogates for the microscale BVP}
\label{Sec:Model}

In classical multiscale formulations such as $\text{FE}^2$ \cite{FE2, FE2_2}, the macroscale constitutive response is obtained by solving a microscale boundary value problem (BVP) at each integration point. The resulting homogenized stresses and consistent tangent moduli are then supplied to the macroscale problem, which itself requires the solution of a nonlinear BVP. As a consequence, the overall computational cost is dominated by the repeated solution of microscale problems throughout the analysis. This renders $\text{FE}^2$ computationally prohibitive in settings such as topology optimization, where the multiscale problem must be solved repeatedly across iterations and design updates.

To overcome this limitation, the microscale boundary value problem can be replaced by a surrogate model that directly approximates the homogenized free energy function $\Psi$. The choice of the surrogate model is restricted by certain thermodynamic, mathematical, and physical conditions \cite{Linden2023-ar}. Neural networks that satisfy these requirements by construction are commonly referred to as Physics-Augmented Neural Networks (PANNs) \cite{rosenkranz2024viscoelasticty,fuhg2024extreme, tepole2025polyconvex}. PANNs rely on specialized architectures with embedded constraints \cite{ICNN, klein2023parametrized, ISNN} to give thermodynamically consistent, physically admissible, and numerically stable representations of the free energy function. As a result of these desirable properties, PANNs have been successfully employed to model complex constitutive behavior across a range of material systems (see e.g. Refs. \cite{klein2022finite, kalina2024magneto, fuhg2024polyconvex, jones2025physics}).

Among these requirements, ellipticity plays a central role in ensuring material stability \cite{zee1983ordinary}. In practice, ellipticity is often ensured via the stronger condition of polyconvexity \cite{alibert1992example, aubert1987counterexample}, which, along with growth coercivity, guarantees the existence of minimizers in nonlinear elasticity \cite{ball1976convexity}. Polyconvexity requires the strain energy function to be convex with respect to the minors of the deformation gradient $\Fb$, i.e., $\Fb$, $\text{Cof}\Fb$, and $\det \Fb$. When adopting an invariant-based representation to enforce objectivity and material symmetry, this requirement translates into nontrivial constraints on the functional dependence of the strain energy. In particular, the invariants must be constructed such that the resulting energy remains convex with respect to the minors of $\Fb$. To ensure this property, the invariants must be constructed such that they are convex functions of $\Fb$ and $\text{Cof}\Fb$. The strain energy function must then be a convex and monotonically non-decreasing function of these invariants \cite{boyd2004convex}. Since the strain energy function takes $J$ directly as input, it only requires convexity with regard to $J$. Furthermore, the strain energy can further be constrained based on the material parameters or microstructural descriptors, in case of prior knowledge on how a given parameter affects the strain energy.

To enforce these constraints, we make use of a modified form of the recently proposed Input Specific Neural Networks (ISNNs) \cite{ISNN}. Their architecture enables us to construct a parametrized strain energy function while enforcing the required physical constraints by design. Below, we summarize the specific network structure used in this study. The network takes as input $\mathbf{x}_0$, $\mathbf{y}_0$, $\mathbf{t}_0$ and predicts the output $\Psi_{\mathcal{NN}}(\mathbf{x}_0$, $\mathbf{y}_0$, $\mathbf{t}_0) \coloneqq x_{H} \in \mathbb{R} $. The inputs and the output are connected through the update formula:
\begin{equation}
    \begin{aligned}
        \mathbf{y}_{h+1} &= \sigma_{mc} \left(  \mathbf{y}_h \mathbf{W}_h^{[yy]^T} + \mathbf{b}_h^{[y]} \right) , \quad &&h = 0, \ldots, H-2 \\
        \mathbf{t}_{h+1} &= \sigma_m \left( \mathbf{t}_h \mathbf{W}_h^{[tt]^T} + \mathbf{b}_h^{[t]} \right) , \quad &&h = 0, \ldots, H-2 \\
        \mathbf{x}_{h+1} &= \sigma_{mc} \left( \mathbf{x}_h \mathbf{W}_h^{[xx]^T} +  \mathbf{x}_0 \mathbf{W}_h^{[xx_{0}]^T} + \mathbf{y}_h \mathbf{W}_h^{[xy]^T}  + \mathbf{t}_h \mathbf{W}_h^{[xt]^T} + \mathbf{b}_h^{[x]} \right) , &&\quad h = 1, \ldots, H - 1
    \end{aligned}
\end{equation}
with
\begin{equation}
\mathbf{x}_{1} = \sigma_{mc} \left( \mathbf{x}_0 \mathbf{W}_0^{[xx]^T} + \mathbf{y}_0 \mathbf{W}_0^{[xy]^T}  + \mathbf{t}_0 \mathbf{W}_0^{[xt]^T} + \mathbf{b}_0^{[x]} \right) \, .
\label{x_eq_2_0}
\end{equation}
These networks are convex with respect to $\mathbf{x}_0$, convex and monotonically nondecreasing with respect to $\mathbf{y}_0$ and only monotonically nondecreasing with respect to $\mathbf{t}_0$ given that $\mathbf{W}_h^{[xx]}$ is non-negative for $h = 1, \ldots, H$, $\mathbf{W}_h^{[yy]}$, $\mathbf{W}_h^{[xy]}$, $\mathbf{W}_h^{[tt]}$ and $\mathbf{W}_h^{[xt]}$ are non-negative for all $h$, $\sigma_{mc}$ is a convex, monotonically non-decreasing function and $\sigma_m$ is a monotonically non-decreasing function. As activation functions, in the following, we use:

\begin{equation}
  \sigma_{mc}(x) = \log(1+\exp (x)) \ , \quad \sigma_m(x)  = \frac{1}{1+ \exp (-x)}  \ ,
\end{equation}

which fulfill the constraint requirements.

For the compressible anisotropic case, the neural network approximates the full free energy function defined in Eq.~\eqref{eq:invariant_psi}. It takes as input $\mathbf{x}_0 = J$ since convexity of the free energy is only required with respect to the volumetric invariant $J$, $\mathbf{y}_0 = \tilde{\bm{\mathcal{I}}}$ where $\tilde{\bm{\mathcal{I}}} = \bm{\mathcal{I}} \setminus \{J\}$ as we need to be convex and monotonically nondecreasing with respect to all other (isotropic and anisotropic) invariants whereas the potential is parametrized via $\mathbf{t}_0 = \phi$ with $\phi$ representing the volume fraction of inclusions in the microstructure and is passed to the monotonically nondecreasing branch of the network. This choice is made since we know the strain energy should increase with increasing volume fraction when the inclusions have a higher modulus than the surrounding matrix material. Although we can also model a parametrized strain energy function which is arbitrary in the volume fraction, it has been shown \cite{ISNN} that imparting the input-output structural relationship into the network helps with the inverse design, particularly in the extrapolation regime, i.e., with volume fractions outside of those included in the training dataset. Moreover, it must be pointed out that using $J$ directly as an input to the free energy potential does not violate objectivity as $\text{det}(\Qb\Fb)=\text{det}\Qb \, \text{det}\Fb=\text{det}\Fb$ since $\text{det}\Qb=1$. 

In contrast, for the nearly incompressible isotropic formulation, the neural network is used only to approximate the isochoric part of the free energy appearing in Eq.~\eqref{eq:psi_mixed}. The volumetric response is handled analytically through the pressure field $\tilde p$. Consequently, the network takes as input only invariants of the isochoric tensor $\bar{\Cb}$ defined in Eq.~\eqref{eq:C_bar}. These inputs are constrained such that the resulting energy is convex and monotonically non-decreasing, ensuring a stable representation of the distortional response consistent with the mixed formulation. Thus, only the $\mathbf{y}$ branch is retained with $\mathbf{y}_0 =\{\bar\Ic_1, \bar\Ic_2 \}$.

Finally, we require the energy potential and the stresses to be zero at the undeformed configuration, which necessitates further modifications to the potential. The stress normalization must also come from the potential itself to ensure thermodynamic consistency. Energy and stress normalization is generally more involved when designing a polyconvex strain energy (see e.g., Refs. \cite{linden2023neural, jadoon2025inverse}) as care must be taken not to violate polyconvexity with the additional modifications to the free energy. However, even if microscale energy may be polyconvex, the homogenized effective energy generally does not admit a polyconvex representation \cite{barchiesi2007loss, braides1994loss}. For numerical stability in the macroscale equilibrium solve, it is nevertheless desirable for the surrogate strain energy to remain convex with respect to the chosen input variables (i.e., the invariants of $\Cb$), as that ensures a positive semidefinite tangent modulus, which leads to a well-conditioned and invertible stiffness operator.

\paragraph{Neural network training}

Here we briefly explain the training of our neural networks, whereas the training results are presented along with the respective microstructures in Section \ref{Sec:NumericalExamples}. Since we rely on structural tensors to model anisotropy, we can also learn the orientation of these structural tensors (see Eqs. \eqref{eq:cubic_ST} and \eqref{eq:TI_ST}) during training. To do so, we employ the Rodrigues' formula to parametrize the rotation tensor $\mathbf{R}$ in terms of a unit vector $\mathbf{p}$ and a rotation angle $\varphi\in [0,2\pi]$ as:
\begin{equation} \label{eq:rodrigues}
    \mathbf{R}(\varphi \pb) = \exp(\varphi \Pb) 
    = \Ib + (\sin \varphi) \Pb + (1-\cos \varphi) \Pb^2 \ ,
\end{equation}
where $\Pb \equiv \varepsilonb \pb$ and $\varepsilonb$ is the third order permutation tensor. Thus, we have these additional trainable parameters on top of the neural network weights and biases. In this work, we assume prior knowledge of the symmetry group associated with the microstructure. However, that might not always be the case. In such a scenario, the anisotropy class itself can also be learned during training \cite{fuhg2022learning, Kalina2025-lq, jadoon2025inverse} along with the preferred direction of anisotropy

Details regarding the generation of training data are provided alongside the corresponding microstructural examples in Section \ref{Sec:NumericalExamples}. For all problems considered in this work, the proposed ISNN architectures are implemented in PyTorch \cite{paszke2019pytorch} and trained using the Adam optimizer \cite{kingma2014adam} for $10^5$ epochs. The full training dataset is processed in a single batch during each optimization step. A learning rate of $10^{-3}$ is used for the neural network weights and biases, while a smaller learning rate of $10^{-4}$ is prescribed for the scalar parameters associated with the rotation tensor. We have found that employing a reduced learning rate for these scalar parameters improves the stability and convergence of the training process.

With the surrogate constitutive models fully defined and trained, we now integrate them into a topology optimization framework. By replacing the microscale boundary value problem with a PANN-based representation of the homogenized free energy, the proposed approach enables efficient evaluation of nonlinear, anisotropic material responses within an optimization loop. This removes the primary computational bottleneck of classical multiscale methods and makes such models tractable for design problems involving repeated forward solves. In the following section, we formulate the topology optimization problem and outline its numerical implementation.

\section{Topology optimization}\label{Sec:TO_problem}

We consider density-based topology optimization of structures undergoing large deformations, solved using a geometrically nonlinear finite element formulation. A total Lagrangian description with reference configuration $\Omega_0$ is adopted. The displacement field and geometry are interpolated using standard isoparametric shape functions. The computational domain $\Omega_0$ is discretized into $N_e$ elements. Each element $e$ is assigned a set of design variables:
\begin{equation}
\bm{\Theta}_e = \{\rho_e, \bm{\mathcal D}_e\},
\end{equation}
where $\rho_e \in [0,1]$ denotes the pseudo-density and $\bm{\mathcal D}_e$ collects additional microstructural design parameters. To suppress checkerboard effects and achieve mesh-independent solutions, scalar design variables are regularized using a linear filter \cite{bourdin2001filters} defined in physical space based on distances between element centroids. While this formulation applies to scalar variables such as density or inclusion volume fraction, orientation variables require a modified filtering approach, which is detailed in Section \ref{Sec:Sens_anisotropic}. Let $\boldsymbol{\xi} \in \mathbb{R}^{N_e}$ denote a vector of elemental design variables. The filtered field is computed as:
\begin{equation}
\tilde{\boldsymbol{\xi}} = \mathbf H\,\boldsymbol{\xi} \oslash \mathbf H_s,
\end{equation}
where $\mathbf H$ is a sparse weighting matrix constructed using a rectifying cone-type kernel :
\begin{equation}\label{eq:Rmin}
H_{ej} = \max\!\left(0,\; r_{\mathrm{min}} - \|\mathbf{x}_e - \mathbf{x}_j\|\right),
\end{equation}
with $\mathbf{x}_e$ denoting the centroid of element $e$ and $r_{\mathrm{min}}$ the filter radius specified in physical units. The vector $\mathbf H_s$ contains the row sums of $\mathbf H$, and $\oslash$ denotes component-wise division. To facilitate binary representation of the pseudo-densities, the filtered density is subsequently projected using a smooth Heaviside function:
\begin{equation}\label{eq:RhoProjection}
\bar{\rho}_e =
\frac{\tanh(\beta_\rho \eta_\rho) + \tanh(\beta_\rho(\tilde{\rho}_e - \eta_\rho))}
{\tanh(\beta_\rho \eta_\rho) + \tanh(\beta_\rho(1-\eta_\rho))},
\end{equation}
where $\beta_\rho$ controls the projection sharpness and $\eta_\rho$ is a threshold parameter. Additional design variables are filtered analogously but are not subjected to projection unless otherwise stated.

In the context of topology optimization, we employ a density-based formulation using the Solid Isotropic Material with Penalization (SIMP) method \cite{SIMP1,SIMP2}. The projected density then defines the interpolation function:
\begin{equation}
\chi_e = \bar{\rho}_e^p(1-\varepsilon) + \varepsilon,
\end{equation}
where $p$ is the penalization exponent and $\varepsilon$ is a small regularization parameter introduced to avoid singular stiffness matrices in near-void regions. Throughout this work, we use $\varepsilon = 10^{-3}$ unless otherwise stated. A major concern when analyzing structures undergoing large deformations is mesh distortion, which can lead to convergence difficulties. This issue is particularly pronounced in density-based topology optimization due to the presence of void-like regions in the design. Various approaches have been proposed in the literature (see e.g. Refs.~\cite{buhl2000stiffness, bruns2003element, yoon2005element}) to mitigate mesh distortion and improve convergence. However, many of these methods rely on linear elastic constitutive assumptions, which may not provide an accurate response for nonlinear hyperelastic materials. To address this issue, we adopt the energy interpolation scheme proposed in Ref.~\cite{wang2014interpolation}, which first introduces a kinematics interpolation strategy in which the deformation gradient is smoothly reduced in low-density regions according to:
\begin{equation}
\hat{\Fb} = \Ib + \gamma(\chi_e)\nabla_0 \ub.
\end{equation}

Here, $\gamma(\chi)$ is a smooth projection function given by:
\begin{equation}\label{eq:gamma_chi}
\gamma(\chi)=
\frac{\tanh(\beta_\psi\eta_\psi)+\tanh(\beta_\psi(\chi-\eta_\psi))}
{\tanh(\beta_\psi\eta_\psi)+\tanh(\beta_\psi(1-\eta_\psi))},
\end{equation}
which ensures a gradual transition between finite-strain and near-linear behavior. The parameters $\beta_\psi$ and $\eta_\psi$ govern the steepness and threshold of the transition, respectively. In this work, we adopt $\beta_\psi = 500$ and $\eta_\psi = 0.01$, consistent with the parameter choices reported in Ref. \cite{wang2014interpolation}, where analogous parameters were shown to yield robust performance across a range of topology optimization problems. While continuation strategies may be employed to further improve convergence, fixed values are used here for simplicity. The corresponding modified right Cauchy-Green tensor and Jacobian are defined as:

\begin{equation}
\hat{\Cb} = \hat{\Fb}^T \hat{\Fb},
\qquad
\hat{J} = \det \hat{\Fb}.
\end{equation}

To further enhance numerical stability, particularly in low-density regions, a linear elastic stabilization term is introduced. The interpolated strain energy for each element is then written in the general form:
\begin{equation}\label{eq:linear_stab}
\tilde{\Psi}_e =
\Psi^{\mathrm{nl}}_e(\hat{\Cb}, \tilde{\bm{\mathcal D}}_e) + w_{lin,e}\,\Psi_{lin}(\varepsilon_e),
\end{equation}
where $\Psi^{\mathrm{nl}}_e(\hat{\Cb}, \tilde{\bm{\mathcal D}}_e)$ denotes the nonlinear strain energy contribution evaluated using the modified kinematics. Its density-based interpolation is specified separately for each constitutive formulation, as it differs between the compressible and nearly incompressible cases. Moreover,
\begin{equation}
w_{lin,e} = \chi_e \bigl(1-\gamma(\chi_e)^2\bigr),
\end{equation}
and $\Psi_{lin}=\frac12\eps_e:\mathbb C_{lin}:\eps_e$ denotes a small-strain linear elastic energy. This interpolation ensures that near-void regions are governed primarily by a linear elastic response, whereas fully solid regions retain the nonlinear constitutive behavior. As the density decreases, the reduction of the kinematic scaling factor $\gamma(\chi_e)$ together with the increasing relative contribution of the linear stabilization term prevents excessive nonlinear deformation in low-stiffness elements. This transition mitigates mesh distortion and element inversion, improves numerical conditioning of the tangent stiffness matrix, and enhances robustness of the Newton iterations.

The objective of topology optimization is to determine an optimal distribution of material and microstructural parameters that minimizes the end compliance \cite{buhl2000stiffness} subject to equilibrium and design constraints. The optimization problem is formulated as

\begin{align}
\label{eq:OptProb}
\text{minimize} \quad & C = \left( \mathbf{f}^{\mathrm{ext}} \right)^T \mathbf{u} \\
\text{subject to} \quad 
& \bm{\mathcal{R}}(\mathbf{q}, \bm{\Theta}) = \mathbf{0}, \notag \\
& g_i(\bm{\Theta}) \leq 0, \quad i = 1, \dots, N_g, \notag \\
& \bm{\Theta}_{\min} \leq \bm{\Theta} \leq \bm{\Theta}_{\max}. \notag
\end{align}

where $C$ denotes the end compliance, $\bm{\mathcal{R}}$ represents the discrete equilibrium residual arising from the finite element formulation, and $\bm{\Theta} = \{\bm{\rho}, \bm{\mathcal{D}}\}$ collects the design variables. Here, $\mathbf{q}$ denotes the vector of state variables arising from the finite element formulation. For the compressible displacement-based formulation, $\mathbf{q}=\mathbf{u}$, whereas for the nearly incompressible mixed formulation, $\mathbf{q} = (\mathbf{u}, \tilde{\mathbf{p}})$ includes both displacement and pressure degrees of freedom. The constraint on the residual $\bm{\mathcal{R}}$ to be zero stems from the global equilibrium, whereas we define two additional constraints as:

\begin{equation}
\begin{aligned}
\label{eq:constraints}
g_1 &= \sum_{e=1}^{N_e} \bar{\rho}_e V_e - \eta V_0 \leq 0, \\
g_2 &= \sum_{e=1}^{N_e} \phi_e \rho_i \bar{\rho}_e V_e - m_{\phi} \leq 0,
\end{aligned}
\end{equation}

Here, $g_1$ enforces a global volume constraint with prescribed volume fraction $\eta \in (0,1)$, while $g_2$ constrains the total mass of inclusions in the microstructure. Depending on the specific problem setting, one or both of these constraints may be active. The latter acts as a cost constraint by limiting the available amount of reinforcing material. In this expression, $\phi_e$ denotes the inclusion volume fraction at the microscale associated with element $e$, $\rho_i$ is the true material density of the inclusions, and $m_{\phi}$ is the prescribed upper bound on the total inclusion mass and is defined as:
\begin{equation}\label{eq:targetMass}
    m_{\phi} = \frac{\rho_i \eta V_0 \phi_{\text{max}}}{c} \ ,
\end{equation}
where $\phi_{\text{max}}$ is the maximum inclusion volume fraction at the microscale, specified with the numerical examples presented in Section \ref{Sec:NumericalExamples} and $c$ is used as a control for the target inclusion mass constraint.

The preceding developments define the common topology optimization framework and the associated optimization problem. The formulation of the equilibrium residual $\bm{\mathcal{R}}$ depends on the choice of constitutive model introduced in Section \ref{Sec:Theory}. In the following, we first consider the compressible anisotropic hyperelastic case, which leads to a displacement-based formulation. The nearly incompressible isotropic case is treated thereafter using a mixed displacement–pressure formulation.

\subsection{Compressible anisotropic hyperelasticity}

For the compressible anisotropic hyperelastic formulation introduced in Section~\ref{Sec:Theory}, the topology optimization problem in Eq.~\eqref{eq:OptProb} leads to a displacement-based finite element discretization. In this case, the state vector reduces to $\mathbf{q} = \mathbf{u}$, and equilibrium is governed by the balance of internal and external forces. We employ standard trilinear hexahedral elements with three displacement degrees of freedom per node. The deformation mapping and displacement field are interpolated using isoparametric shape functions. The governing equilibrium equations follow from the stationarity of the total potential energy functional:

\begin{equation}
\Pi(\mathbf{u}) = 
\int_{\Omega_0}\!{\tilde\Psi}\,\mathrm{d}\Omega_0
- \int_{\Omega_0}\!\mathbf{b}_0^{\top}\mathbf{u}\,\mathrm{d}\Omega_0
- \int_{\Gamma_0^t}\!\bar{\mathbf{t}}^{\top}\mathbf{u}\,\mathrm{d}\Gamma_0,
\end{equation}

which is consistent with the strong and weak forms introduced in Section~\ref{Sec:Theory}. Here, $\tilde\Psi$ is the modified strain energy from Eq. \eqref{eq:linear_stab}. Enforcing $\delta \Pi = 0$ yields the discrete residual equation:

\begin{equation}
\bm{\mathcal{R}}(\mathbf{u}) = \mathbf{f}_\text{int}(\mathbf{u}, \bm{\Theta}) - \mathbf{f}_\text{ext} = \mathbf{0}.
\end{equation}

The external force vector collects contributions from body forces and surface tractions:

\begin{equation}
\mathbf{f}_\text{ext} =
\int_{\Omega_0} \mathbf{N}^\top \mathbf{b}_0\,\mathrm{d}\Omega_0
+ \int_{\Gamma_0^t} \mathbf{N}^\top \bar{\mathbf{t}}\,\mathrm{d}\Gamma_0.
\end{equation}

The internal force vector is assembled element-wise and consists of contributions from the nonlinear hyperelastic response and the linear stabilization term introduced in Section~\ref{Sec:TO_problem}. Using the energy decomposition $\tilde{\Psi}_e = \Psi^{\mathrm{nl}}_e + w_{lin,e}\Psi_{lin}$, the internal force vector reads:
\begin{equation}
\mathbf f_\text{int}
=
\sum_{e=1}^{N_e}
\int_{\Omega_0^e}
\Big[
\gamma(\chi_e)\,\mathbf B_N^{\top}\mathbf S
+
\mathbf B_{N}^{\mathrm{lin}\top}
\bigl(
w_{lin,e}\,\mathbf S_{lin}
\bigr)
\Big]
\,\mathrm d\Omega_0 ,
\end{equation}
where $\mathbf B_N$ denotes the total-Lagrangian strain-displacement matrix associated with the second Piola-Kirchhoff stress $\mathbf S$, and $\mathbf B_N^{\mathrm{lin}}$ its small-strain counterpart.

For the compressible case, the nonlinear strain energy contribution is defined as:
\begin{equation}
\Psi^{\mathrm{nl}}_e = \chi_e\,\Psi(\hat{\Cb}_e, \tilde{\bm{\mathcal D}}_e)
= \chi_e\,\Psi(\hat{\bm{\mathcal I}}_e, \tilde{\bm{\mathcal D}}_e),
\end{equation}
where $\hat{\bm{\mathcal I}}$ denotes the set of invariants associated with the modified right Cauchy--Green tensor $\hat{\Cb}$. These invariants are defined analogously to those introduced in Section~\ref{Sec:Theory}, but are evaluated using $\hat{\Cb}$ in place of $\Cb$. This formulation ensures that the constitutive response is evaluated consistently with the modified kinematics introduced for numerical stabilization. The second Piola-Kirchhoff stress follows as:
\begin{equation}
\Sb = 2 \frac{\partial \Psi^{\mathrm{nl}}_e}{\partial \hat{\Cb}}
= 2 \chi_e \sum_i \frac{\partial \Psi}{\partial \hat{\Ic}_i} \frac{\partial \hat{\Ic}_i}{\partial \hat{\Cb}} .
\end{equation}

The consistent tangent matrix $\mathbf K_T = \partial \bm{\mathcal{R}} / \partial \mathbf{u}$ is assembled from element contributions. Specifically, at each Gauss point:
\begin{equation}
\mathbf k_{e}^{(g)}
=
\gamma(\chi_e)^2
\Bigl(
\underbrace{\mathbf B_N^\top\,\mathbb C_\mathrm{alg}\,\mathbf B_N}_{\text{material}}
+
\underbrace{\mathbf B_G^\top\,\mathbf S_\mathrm{head}\,\mathbf B_G}_{\text{geometric}}
\Bigr)
+
\underbrace{\mathbf B_{N}^{\mathrm{lin}\top}\,\bigl(w_{\mathrm{lin},e}\,\mathbb C_\mathrm{lin}\bigr)\,\mathbf B_N^\mathrm{lin}}_{\text{stabilization}},
\label{eq:ke_elem}
\end{equation}
and
\begin{equation}
\mathbf k_e=\sum_{g=1}^{8} w_g\,J_g\,\mathbf k_{e}^{(g)}.
\end{equation}
Here $w_g$ and $J_g$ denote the Gauss quadrature weights and the determinant of the element Jacobian at the integration point $g$, respectively. Furthermore, $\mathbf B_G$ is the standard total-Lagrangian geometric matrix and $\mathbf S_\mathrm{head}$ is the block-diagonal stress matrix constructed from the second Piola-Kirchhoff stress evaluated under the modified kinematics:

\begin{equation}
\mathbf S_\mathrm{head}=\mathrm{diag}(\mathbf S,\mathbf S,\mathbf S).
\end{equation}

The consistent algorithmic tangent is obtained from the nonlinear strain energy contribution as:
\begin{equation}
\mathbb{C}_\mathrm{alg}
=
4 \chi_e \sum_i \sum_j \left[
\frac{\partial^2 \Psi}{\partial \hat{\Ic}_i \partial \hat{\Ic}_j}
\left( \partial_{\hat{\Cb}} \hat{\Ic}_i \otimes \partial_{\hat{\Cb}} \hat{\Ic}_j \right)
+
\frac{\partial \Psi}{\partial \hat{\Ic}_i}
\left( \partial_{\hat{\Cb}^2} \hat{\Ic}_i \right)
\right].
\label{eq:C_alg}
\end{equation}

While the displacement-based formulation presented above is suitable for compressible materials, it becomes inadequate in the nearly incompressible regime. In such cases, the volumetric response is dominated by a high bulk modulus, which leads to numerical difficulties such as volumetric locking when using standard displacement-based finite elements.

To address this issue, we adopt a mixed displacement-pressure formulation, in which an additional pressure field is introduced to enforce the incompressibility constraint. This formulation is consistent with the constitutive framework introduced in Section \ref{Sec:Theory}, where the strain energy is decomposed into isochoric and volumetric contributions. In contrast to the compressible case, we restrict attention here to isotropic material behavior and only consider topology as the design variable, without additional microscale parameters. The corresponding finite element formulation and discretized system are presented in the following subsection.

\subsection{Nearly incompressible isotropic hyperelasticity}

For nearly incompressible materials, the displacement-based formulation becomes inadequate due to volumetric locking. To overcome this limitation, we adopt a mixed displacement-pressure finite element formulation. In this case, the state vector is augmented as:

\begin{equation}
\mathbf q = \{\mathbf u, \tilde{\mathbf p}\},
\end{equation}
where $\tilde{\mathbf p}$ denotes the elemental pressure degrees of freedom. The displacement field is interpolated using standard trilinear hexahedral shape functions, while the pressure field is approximated using a single constant degree of freedom per element (i.e., a $P_0$ approximation). The governing equations follow from the stationarity of the mixed potential functional:

\begin{equation}
\Pi(\mathbf u,\tilde{\mathbf p}) =
\int_{\Omega_0} \tilde{\Psi}(\hat{\Cb},\tilde{\mathbf p})\,\mathrm d\Omega_0
- \int_{\Omega_0} \mathbf b_0^\top \mathbf u\,\mathrm d\Omega_0
- \int_{\Gamma_0^t} \bar{\mathbf t}^\top \mathbf u\,\mathrm d\Gamma_0,
\end{equation}
which is consistent with the strong and weak forms introduced in Section~\ref{Sec:Theory}. Taking variations with respect to $\mathbf u$ and $\tilde{\mathbf p}$ yields the coupled residual equations (see Eq. \eqref{eq:weak_mixed}):
\begin{equation}
\bm{\mathcal R}_u(\mathbf u,\tilde{\mathbf p}) = \mathbf 0,
\qquad
\bm{\mathcal R}_p(\mathbf u,\tilde{\mathbf p}) = \mathbf 0.
\end{equation}

The modified deformation gradient and right Cauchy-Green tensor are defined analogously to the compressible case as:
\begin{equation}
\hat{\Fb} = \Ib + \gamma(\chi_e)\nabla_0 \mathbf u,
\qquad
\hat{\Cb} = \hat{\Fb}^\top \hat{\Fb},
\qquad
\hat{J} = \det(\hat{\Fb}),
\end{equation}
and the isochoric part of the deformation is given by:
\begin{equation}
\bar{\hat{\Cb}} = \hat{J}^{-2/3}\hat{\Cb}.
\end{equation}

Consistent with the constitutive framework introduced in Section \ref{Sec:Theory}, the strain energy is decomposed into isochoric and volumetric contributions. The resulting nonlinear strain energy at the element level is defined as:

\begin{equation}
\Psi^{\mathrm{nl}}_e
=
\chi_e\,\Psi_{\mathrm{iso}}(\bar{\hat{\Cb}}_e)
-
\tilde{p}_e(\hat{J}_e-1)
-
\frac{1}{2\kappa_e}\tilde{p}_e^2,
\end{equation}
where $\Psi_{\mathrm{iso}}$ denotes the isochoric free energy, and $\kappa_e = \chi_e \kappa_0$ is the density-dependent bulk modulus. The second Piola-Kirchhoff stress follows as:

\begin{equation}
\Sb
=
2 \frac{\partial \Psi^{\mathrm{nl}}_e}{\partial \hat{\Cb}}
=
\chi_e\,\mathbf S_{\mathrm{iso}}(\bar{\hat{\Cb}})
-
\tilde{p}_e \hat{J}\hat{\Cb}^{-1},
\end{equation}
where $\mathbf S_{\mathrm{iso}} = 2\partial \Psi_{\mathrm{iso}} / \partial {\hat{\Cb}}$ denotes the isochoric stress contribution derived from $\Psi_{\mathrm{iso}}$ through the chain rule. The displacement residual retains the same stabilized structure as in the compressible case, i.e.,

\begin{equation}
\mathbf f_{\text{int},u}
=
\sum_{e=1}^{N_e}
\int_{\Omega_0^e}
\Big[
\gamma(\chi_e)\,\mathbf B_N^\top \mathbf S
+
\mathbf B_{N}^{\mathrm{lin}\top}
\bigl(
w_{lin,e}\,\mathbf S_{lin}
\bigr)
\Big]
\,\mathrm d\Omega_0,
\end{equation}
where $w_{lin,e} = \chi_e\bigl(1-\gamma(\chi_e)^2\bigr)$. The pressure residual enforces the incompressibility constraint in weak form:

\begin{equation}
R_p^e
=
\int_{\Omega_0^e}
\left[
-(\hat{J}-1) - \frac{\tilde{p}_e}{\kappa_e}
\right]
\,\mathrm d\Omega_0,
\end{equation}
reflecting the constraint introduced in Section~\ref{Sec:Theory}. The linearized system results in the block structure:

\begin{equation}
\mathbf K_e =
\begin{bmatrix}
\mathbf K_{uu} & \mathbf K_{u p} \\
\mathbf K_{p u} & K_{p p}
\end{bmatrix},
\end{equation}
where the displacement block retains the same stabilized structure as in the compressible case:
\begin{equation}
\mathbf K_{uu}^{(g)}
=
\gamma(\chi_e)^2
\Bigl(
\mathbf B_N^\top\,\mathbb C_\mathrm{alg}\,\mathbf B_N
+
\mathbf B_G^\top\,\mathbf S_\mathrm{head}\,\mathbf B_G
\Bigr)
+
\mathbf B_{N}^{\mathrm{lin}\top}
\bigl(
w_{lin,e}\,\mathbb C_\mathrm{lin}
\bigr)
\mathbf B_N^\mathrm{lin},
\end{equation}

whereas the displacement-pressure coupling blocks follow from the linearization of the mixed residuals:
\begin{equation}
\mathbf K_{u p}
=
\int_{\Omega_0^e}
\gamma(\chi_e)\,\mathbf B_N^\top
\left(
-\hat{J}\hat{\Cb}^{-1}
\right)
\,\mathrm d\Omega_0,
\end{equation}
\begin{equation}
\mathbf K_{p u}
=
\int_{\Omega_0^e}
\left(
-\frac{\partial \hat{J}}{\partial \mathbf u}
\right)
\,\mathrm d\Omega_0,
\end{equation}
and
\begin{equation}
K_{p p}
=
\int_{\Omega_0^e}
\left(
-\frac{1}{\kappa_e}
\right)
\,\mathrm d\Omega_0.
\end{equation}

The linear stabilization term contributes only to the displacement block $\mathbf K_{uu}$, since it is independent of the pressure variable. The coupling blocks $\mathbf K_{up}$ and $\mathbf K_{pu}$ arise solely from the mixed nonlinear contribution. For a consistent linearization of the underlying mixed potential, these blocks are transpose-equivalent at the variational level, although they are expressed in different forms in the implementation.

\subsection{Sensitivity analysis}\label{sec:sensitivity}

The topology optimization problem is solved using a gradient-based approach, for which sensitivities of the compliance objective with respect to the design variables are evaluated via the adjoint method. Let the objective be denoted by:
\begin{equation}
C = \left(\mathbf f^{\mathrm{ext}}\right)^{\top}\mathbf u,
\end{equation}
and let $\bm{\mathcal R}(\mathbf q,\bm{\Theta})=\mathbf 0$ denote the discrete equilibrium equations introduced in Eq.~\eqref{eq:OptProb}. Here, $\mathbf q$ is the state vector, which reduces to $\mathbf q=\mathbf u$ in the compressible case and is augmented as $\mathbf q=\{\mathbf u,\tilde{\mathbf p}\}$ in the nearly incompressible case.

For a generic elemental design variable $\Theta_e$, the total derivative of the compliance reads:
\begin{equation}
\frac{\partial C}{\partial \Theta_e}
=
\frac{\partial C}{\partial \mathbf q}
\frac{\partial \mathbf q}{\partial \Theta_e}.
\end{equation}
Direct evaluation of $\partial \mathbf q/\partial \Theta_e$ is avoided by introducing the augmented objective:
\begin{equation}
\hat C
=
C
+
\bm{\lambda}^{\top}\bm{\mathcal R}(\mathbf q,\bm{\Theta}),
\end{equation}
where $\bm{\lambda}$ is the adjoint variable. Since $\bm{\mathcal R}=\mathbf 0$ at equilibrium, one has $\hat C=C$. Differentiating with respect to $\Theta_e$ yields:
\begin{equation}
\frac{\partial \hat C}{\partial \Theta_e}
=
\frac{\partial C}{\partial \mathbf q}
\frac{\partial \mathbf q}{\partial \Theta_e}
+
\bm{\lambda}^{\top}
\left(
\frac{\partial \bm{\mathcal R}}{\partial \mathbf q}
\frac{\partial \mathbf q}{\partial \Theta_e}
+
\frac{\partial \bm{\mathcal R}}{\partial \Theta_e}
\right).
\end{equation}
Rearranging gives:
\begin{equation}
\frac{\partial \hat C}{\partial \Theta_e}
=
\left(
\frac{\partial C}{\partial \mathbf q}
+
\bm{\lambda}^{\top}\frac{\partial \bm{\mathcal R}}{\partial \mathbf q}
\right)
\frac{\partial \mathbf q}{\partial \Theta_e}
+
\bm{\lambda}^{\top}
\frac{\partial \bm{\mathcal R}}{\partial \Theta_e}.
\end{equation}
Choosing the adjoint vector such that:
\begin{equation}
\left(\frac{\partial \bm{\mathcal R}}{\partial \mathbf q}\right)^{\top}\bm{\lambda}
=
-
\left(\frac{\partial C}{\partial \mathbf q}\right)^{\top},
\label{eq:adjoint_problem}
\end{equation}
eliminates the dependence on the state sensitivities, and the objective gradient reduces to:
\begin{equation}
\frac{\partial C}{\partial \Theta_e}
=
\bm{\lambda}^{\top}
\frac{\partial \bm{\mathcal R}}{\partial \Theta_e}.
\label{eq:adjoint_gradient_general}
\end{equation}

For the compressible formulation, $\mathbf q=\mathbf u$ and
\begin{equation}
\frac{\partial C}{\partial \mathbf q}
=
\frac{\partial C}{\partial \mathbf u}
=
\mathbf f^{\mathrm{ext}},
\end{equation}
such that Eq.~\eqref{eq:adjoint_problem} reduces to:
\begin{equation}
\mathbf K_T^{\top}\bm{\lambda}
=
-
\mathbf f^{\mathrm{ext}}.
\end{equation}

For the nearly incompressible mixed formulation, the residual consists of displacement and pressure contributions:
\begin{equation}
\bm{\mathcal R}(\mathbf q,\bm{\Theta})
=
\begin{bmatrix}
\bm{\mathcal R}_u \\
\bm{\mathcal R}_p
\end{bmatrix},
\end{equation}
and the compliance depends only on the displacement field. Therefore,
\begin{equation}
\frac{\partial C}{\partial \mathbf q}
=
\begin{bmatrix}
\mathbf f^{\mathrm{ext}} \\
\mathbf 0
\end{bmatrix},
\end{equation}
and the adjoint problem becomes:
\begin{equation}
\begin{bmatrix}
\mathbf K_{uu} & \mathbf K_{up} \\
\mathbf K_{pu} & K_{pp}
\end{bmatrix}^{\top}
\begin{bmatrix}
\bm{\lambda}_u \\
\lambda_p
\end{bmatrix}
=
-
\begin{bmatrix}
\mathbf f^{\mathrm{ext}} \\
\mathbf 0
\end{bmatrix}.
\label{eq:adjoint_mixed}
\end{equation}

It is important to note that the design variables enter the finite element formulation through filtering and projection outlined in Section \ref{Sec:TO_problem}. In particular, the density dependence follows the chain:
\begin{equation}
\rho_e \;\rightarrow\; \tilde{\rho}_e \;\rightarrow\; \bar{\rho}_e \;\rightarrow\; \chi_e,
\end{equation}
whereas additional microstructural variables, such as $\phi_e$ and $\theta_e$, are filtered but not projected:
\begin{equation}
\phi_e \;\rightarrow\; \tilde{\phi}_e,
\qquad
\theta_e \;\rightarrow\; \tilde{\theta}_e.
\end{equation}
Accordingly, the objective sensitivities with respect to the raw design variables are obtained using the chain rule:
\begin{equation}
\frac{\partial C}{\partial \rho_e}
=
\frac{\partial C}{\partial \bar{\rho}_e}
\frac{\partial \bar{\rho}_e}{\partial \tilde{\rho}_e}
\frac{\partial \tilde{\rho}_e}{\partial \rho_e},
\label{eq:rho_chain_rule}
\end{equation}
\begin{equation}
\frac{\partial C}{\partial \phi_e}
=
\frac{\partial C}{\partial \tilde{\phi}_e}
\frac{\partial \tilde{\phi}_e}{\partial \phi_e},
\qquad
\frac{\partial C}{\partial \theta_e}
=
\frac{\partial C}{\partial \tilde{\theta}_e}
\frac{\partial \tilde{\theta}_e}{\partial \theta_e}.
\label{eq:phi_theta_chain_rule}
\end{equation}
The element sensitivity kernels therefore evaluate derivatives with respect to the processed variables \(\bar{\rho}_e\), \(\tilde{\phi}_e\), and \(\tilde{\theta}_e\), after which Eqs.~\eqref{eq:rho_chain_rule}-\eqref{eq:phi_theta_chain_rule} are used to recover sensitivities with respect to the raw design variables. 

\subsubsection{Compressible anisotropic case}\label{Sec:Sens_anisotropic}

For the compressible formulation, the residual depends on the design variables through the interpolated constitutive scaling $\chi_e$, the kinematic interpolation $\gamma(\chi_e)$, the invariant-based constitutive response, and the linear stabilization term. At the element level, the derivative of the displacement residual with respect to the projected density $\bar{\rho}_e$ is written as
\begin{align}
\frac{\partial \mathbf R_u}{\partial \bar{\rho}_e}
&=
\underbrace{
\frac{\partial (\chi_e \gamma_e)}{\partial \bar{\rho}_e}
\,\mathbf B_N^{\top}\mathbf S
}_{\text{interpolation}}
+
\underbrace{
\chi_e \gamma_e
\left(
\frac{\partial \mathbf B_N}{\partial \bar{\rho}_e}
\right)^{\top}
\mathbf S
}_{\text{kinematic}}
\nonumber\\
&\quad+
\underbrace{
\chi_e \gamma_e
\mathbf B_N^{\top}
\frac{\partial \mathbf S}{\partial \bar{\rho}_e}
}_{\text{constitutive}}
+
\underbrace{
\mathbf B_N^{\mathrm{lin}\top}
\left(
\frac{\partial w_{lin,e}}{\partial \bar{\rho}_e}
\mathbf S_{lin}
\right)
}_{\text{stabilization}}.
\label{eq:dR_drho_comp}
\end{align}

The constitutive contribution follows from the invariant-based free energy representation as:
\begin{equation}
\frac{\partial \mathbf S}{\partial \bar{\rho}_e}
=
2\frac{\partial \chi_e}{\partial \bar{\rho}_e}
\sum_i
\frac{\partial \Psi}{\partial \hat{\Ic}_i}
\frac{\partial \hat{\Ic}_i}{\partial \hat{\Cb}}.
\label{eq:dS_drho_comp}
\end{equation}

\paragraph{Microstructural design variables}

For additional microstructural design variables, the dependence enters only through the constitutive response. In the cubic RVE setting considered in Section~\ref{sec:cubicRVE}, the additional design variable is the local inclusion volume fraction $\phi_e$. In that case, the stress derivative with respect to the filtered variable $\tilde{\phi}_e$ is:
\begin{equation}
\frac{\partial \mathbf S}{\partial \tilde{\phi}_e}
=
2\chi_e
\sum_i
\frac{\partial^2 \Psi}{\partial \hat{\Ic}_i\,\partial \tilde{\phi}_e}
\hat{\mathbf B}_i,
\label{eq:dS_dphi}
\end{equation}
where
\begin{equation}
\hat{\mathbf B}_i = \frac{\partial \hat{\Ic}_i}{\partial \hat{\Cb}}.
\end{equation}
At the residual level, this yields:
\begin{equation}
\frac{\partial \mathbf R_u}{\partial \tilde{\phi}_e}
=
\chi_e\gamma_e
\mathbf B_N^{\top}
\frac{\partial \mathbf S}{\partial \tilde{\phi}_e}.
\label{eq:dR_dphi}
\end{equation}

\paragraph{Fiber orientation sensitivity}

For the fiber-reinforced transversely isotropic RVE, the set of design variables is $\bm{\Theta}=\{\bm{\rho},\bm{\phi},\bm{\theta}\}$, where $\theta_e$ parametrizes the fiber orientation in a prescribed plane. In this case, the invariants and their associated basis tensors depend explicitly on the filtered orientation variable $\tilde{\theta}_e$. The resulting stress sensitivity is:
\begin{equation}
\frac{\partial \mathbf S}{\partial \tilde{\theta}_e}
=
2\chi_e
\sum_i
\left(
\sum_j
\frac{\partial^2 \Psi}{\partial \hat{\Ic}_i \partial \hat{\Ic}_j}
\frac{\partial \hat{\Ic}_j}{\partial \tilde{\theta}_e}
\hat{\mathbf B}_i
+
\frac{\partial \Psi}{\partial \hat{\Ic}_i}
\frac{\partial \hat{\mathbf B}_i}{\partial \tilde{\theta}_e}
\right).
\label{eq:dS_dtheta}
\end{equation}

The corresponding residual derivative is:
\begin{equation}
\frac{\partial \mathbf R_u}{\partial \tilde{\theta}_e}
=
\chi_e\gamma_e
\mathbf B_N^{\top}
\frac{\partial \mathbf S}{\partial \tilde{\theta}_e}.
\label{eq:dR_dtheta}
\end{equation}

It is important to note that the microstructural design variables $\phi_e$ and $\theta_e$ do not influence the deformation mapping or the interpolated kinematics. In particular, the modified deformation gradient $\hat{\Fb}$ depends only on the displacement field and the density-based interpolation through $\gamma(\chi_e)$, but is independent of $\phi_e$ and $\theta_e$. As a result, the strain-displacement matrices $\mathbf B_N$ and $\mathbf B_N^{\mathrm{lin}}$ remain unaffected by these variables. Consequently, the sensitivities with respect to $\phi_e$ and $\theta_e$ arise solely through the constitutive response, as reflected in Eqs.~\eqref{eq:dS_dphi}--\eqref{eq:dS_dtheta}. It is reiterated here that the fiber orientation does not need to be fed to ISNN as an additional input parameter. It is rather an added advantage of the invariant-based formulation that once trained for a given preferred direction of anisotropy, it can represent the free energy for any preferred direction of the given anisotropy. 

In contrast to the other design variables, the preferred in-plane fiber orientation $\theta$ requires special treatment during filtering due to its periodic nature. In particular, orientations separated by $\pi$ represent the same physical direction, i.e., $\theta \equiv \theta + \pi$. As a result, $\theta$ does not lie on a linear space, and direct application of standard density filtering leads to non-physical interpolation artifacts. To address this, we map the orientation variable $\boldsymbol{\theta} \in [0,1]^{N_e}$ to a doubled angular representation:
\begin{equation}
\boldsymbol{\vartheta} = 2\pi \boldsymbol{\theta},
\end{equation}
and embed it onto the unit circle via:
\begin{equation}
\mathbf{c} = \cos(\boldsymbol{\vartheta}), 
\quad 
\mathbf{s} = \sin(\boldsymbol{\vartheta}),
\end{equation}
where the trigonometric functions are applied component-wise. Importantly, the filtering operation is performed in the embedded $(\cos\vartheta, \sin\vartheta)$ space rather than directly on $\theta$, ensuring that equivalent orientations are treated consistently during interpolation.

A weighted filtering operation is then applied in analogy with the scalar filter. Let $\mathbf{w} \in \mathbb{R}^{N_e}$ denote a vector of element-wise weights defined as:
\begin{equation}
\mathbf{w} = \boldsymbol{\rho}^{\text{proj}} \odot \boldsymbol{\alpha}^{\text{filt}},
\end{equation}
where $\odot$ denotes component-wise multiplication. The filtered cosine and sine fields are given by:
\begin{equation}
\tilde{\mathbf{c}} = \mathbf H\,(\mathbf{w} \odot \mathbf{c}) \oslash (\mathbf H\,\mathbf{w}), 
\quad
\tilde{\mathbf{s}} = \mathbf H\,(\mathbf{w} \odot \mathbf{s}) \oslash (\mathbf H\,\mathbf{w}).
\end{equation}

To ensure consistency on the unit circle, the filtered vectors are renormalized as:
\begin{equation}
\tilde{\mathbf{c}} \leftarrow \frac{\tilde{\mathbf{c}}}{\sqrt{\tilde{\mathbf{c}}^2 + \tilde{\mathbf{s}}^2}}, 
\quad
\tilde{\mathbf{s}} \leftarrow \frac{\tilde{\mathbf{s}}}{\sqrt{\tilde{\mathbf{c}}^2 + \tilde{\mathbf{s}}^2}},
\end{equation}
where all operations are performed component-wise. Finally, the filtered orientation field is recovered via:

\begin{equation}
\tilde{\boldsymbol{\theta}} = \frac{1}{2\pi} \operatorname{atan2}(\tilde{\mathbf{s}}, \tilde{\mathbf{c}}).
\end{equation}

This approach avoids spurious averaging effects associated with direct interpolation of angular quantities. In particular, orientations that are equivalent on the unit circle (e.g., $0^\circ$ and $180^\circ$) map to the same point in $(\cos\vartheta,\sin\vartheta)$ space, and are therefore interpolated consistently. Sensitivities with respect to $\theta$ are computed through this mapping using the chain rule, accounting for the dependence of $(\hat{c}, \hat{s})$ on the filtered design variables. In addition, this filtering strategy promotes smooth spatial variation of the orientation field across neighboring elements. As a result, the preferred directions form a continuous vector field at the macroscale, which can be physically interpreted as coherent fiber paths.

\subsubsection{Nearly incompressible case}

For the mixed nearly incompressible formulation, only the density field is optimized, and the sensitivity must account for both displacement and pressure residuals. Using Eq.~\eqref{eq:adjoint_gradient_general}, the compliance sensitivity is written as:
\begin{equation}
\frac{\partial C}{\partial \bar{\rho}_e}
=
\bm{\lambda}_u^{\top}
\frac{\partial \mathbf R_u}{\partial \bar{\rho}_e}
+
\lambda_p
\frac{\partial R_p}{\partial \bar{\rho}_e}.
\label{eq:adjoint_incomp}
\end{equation}

The derivative of the displacement residual retains the same structure as in the compressible case,
\begin{align}
\frac{\partial \mathbf R_u}{\partial \bar{\rho}_e}
&=
\frac{\partial \gamma_e}{\partial \bar{\rho}_e}
\mathbf B_N^{\top}\mathbf S
+
\gamma_e
\left(
\frac{\partial \mathbf B_N}{\partial \bar{\rho}_e}
\right)^{\top}
\mathbf S
\nonumber\\
&\quad+
\gamma_e
\mathbf B_N^{\top}
\frac{\partial \mathbf S}{\partial \bar{\rho}_e}
+
\mathbf B_N^{\mathrm{lin}\top}
\left(
\frac{\partial w_{lin,e}}{\partial \bar{\rho}_e}
\mathbf S_{lin}
\right),
\label{eq:dRu_drho_incomp}
\end{align}

which again matches the structure implemented in the incompressible element sensitivity kernel.

The corresponding stress derivative contains both isochoric and volumetric contributions:
\begin{equation}
\frac{\partial \mathbf S}{\partial \bar{\rho}_e}
=
\frac{\partial \chi_e}{\partial \bar{\rho}_e}\,\mathbf S_{\mathrm{iso}}
+
\chi_e
\frac{\partial \mathbf S_{\mathrm{iso}}}{\partial \bar{\rho}_e}
-
\tilde{p}_e
\left(
\frac{\partial \hat{J}}{\partial \bar{\rho}_e}\hat{\Cb}^{-1}
+
\hat{J}\frac{\partial \hat{\Cb}^{-1}}{\partial \bar{\rho}_e}
\right).
\label{eq:dS_drho_incomp}
\end{equation}

The pressure residual sensitivity follows directly from the weak incompressibility constraint:
\begin{equation}
\frac{\partial R_p}{\partial \bar{\rho}_e}
=
\int_{\Omega_0^e}
\left[
-\frac{\partial \hat{J}}{\partial \bar{\rho}_e}
+
\frac{\tilde{p}_e}{\kappa_e^2}
\frac{\partial \kappa_e}{\partial \bar{\rho}_e}
\right]
\,\mathrm d\Omega_0.
\label{eq:dRp_drho}
\end{equation}
Since $\kappa_e=\chi_e\kappa_0$, one has:
\begin{equation}
\frac{\partial \kappa_e}{\partial \bar{\rho}_e}
=
\kappa_0
\frac{\partial \chi_e}{\partial \bar{\rho}_e}.
\end{equation}

The analytical sensitivities derived above are verified against finite-difference approximations in Appendix~\ref{app:adjoint_verification}. In particular, the appendix reports finite-difference checks for $\rho_e$, $\phi_e$, and $\theta_e$ in the compressible setting, and for $\rho_e$ in the nearly incompressible formulation, demonstrating excellent agreement. Importantly, the sensitivities required for the optimization are not obtained via automatic differentiation. Instead, analytical expressions for the first and second derivatives of the ISNN-based surrogate, as derived in Ref. \cite{ISNN}, are used to evaluate stresses and consistent tangent moduli. This approach ensures computational efficiency and numerical robustness, and avoids the computational overhead associated with automatic differentiation in nonlinear finite element analyses.

In the present work, the topology optimization problem is solved using the Method of Moving Asymptotes (MMA) \cite{svanberg1987method}, which is widely adopted in topology optimization due to its robustness and efficiency for large-scale problems with a high number of design variables \cite{bendsoe2013topology}. In such settings, second-order optimization methods would require the evaluation and storage of Hessian information, which is computationally prohibitive. MMA, on the other hand, provides a first-order gradient-based alternative by constructing a sequence of separable convex subproblems. The sensitivities entering MMA are obtained after consistent backpropagation through the filtering and projection operations described earlier, ensuring that the optimization is performed with respect to the original design variables. The resulting framework enables the simultaneous optimization of the macroscale density field and, in the compressible case, the microscale parameters governing the anisotropic response. In the following section, we demonstrate the performance of the proposed formulation through a series of numerical examples, highlighting both the accuracy of the sensitivity implementation and the effectiveness of the resulting optimized designs.

\section{Numerical examples}\label{Sec:NumericalExamples}

In this section, we demonstrate the proposed framework on a series of numerical examples. For all the examples presented in this section, we choose $\beta_\psi = 500$ and $\eta_\psi = 0.01$ \cite{wang2014interpolation} in Eq. \eqref{eq:gamma_chi}, and $\eta_\rho=0.5$ in Eq. \eqref{eq:RhoProjection} \cite{bessa2025}. Moreover, a continuation scheme is used on the SIMP penalty exponent $p$ and the projection sharpness parameter $\beta_{\rho}$ which follows directly from Ref. \cite{bessa2025}, i.e., both variables are initialized as $p=\beta_\rho=1$ and the SIMP penalty exponent is increased in increments of $\Delta p$ to a maximum value of $p=4$. Once this maximum value for the penalty exponent is reached, the projection sharpness parameter $\beta_\rho$ is doubled in each continuation step to a maximum value of $\beta_\rho = 8$. The minimum number of iterations in a continuation step is chosen to be 20 whereas the maximum optimization iterations in a continuation step are set to 50. The next continuation step therefore begins when the maximum iterations are reached or when the difference in the objective over the previous five iterations falls below a tolerance of $0.1\%$. For all the examples presented here, the parameters entering the linear stabilization term in Eq.~\eqref{eq:linear_stab} are chosen to be small in magnitude so that they act purely as numerical regularization. In low-density regions, the modified interpolation effectively replaces the finite-strain response with a small-strain linear elastic behavior of very low stiffness, which helps prevent mesh distortion and improves numerical stability. Accordingly, $E=0.01$ MPa and $\nu=0.3$ are used to define the Lam\'e-type constants in Eq.~\eqref{eq:linear_stab}, and should not be interpreted as physical material parameters. Furthermore, all the inclusions in the microstructure are assumed to have a density $\rho_i=2.1\ g/cm^3$. All the other variables that are problem-specific are given with the respective examples.

For all examples presented here, the nonlinear equilibrium equations are solved using a load-controlled Newton-Raphson scheme. A single load step is used whenever convergence is achieved. Otherwise, the load is subdivided and applied incrementally until convergence is recovered, yielding an adaptive load-stepping strategy. The linear systems arising in each Newton iteration are solved using a direct sparse solver, and convergence of the nonlinear iterations is enforced using a prescribed residual tolerance. Since the material response is path-independent and the objective is evaluated at the final converged configuration, the sensitivities are computed from the converged state at the final load step. In the following, we present results for both compressible and nearly incompressible material settings. We begin with the compressible case, followed by examples demonstrating the performance of the proposed framework in the nearly incompressible regime. To validate the proposed framework, Appendix~\ref{App:PhenvsPANNvalidation} presents single-scale topology optimization results for both compressible and nearly incompressible isotropic hyperelastic materials, where PANN-based constitutive models are compared against their corresponding phenomenological models, demonstrating excellent agreement in both optimized topologies and convergence behavior.

\subsection{Compressible case}
We consider two distinct representative volume elements (RVEs) corresponding to different anisotropy classes in order to demonstrate the generality of the proposed framework. The first RVE consists of a fiber-reinforced microstructure exhibiting transverse isotropy, while the second RVE comprises a matrix with a single spherical inclusion, leading to cubic anisotropy. For both cases, optimization is performed concurrently across scales, where the macroscale material distribution is controlled through the density field $\rho$, and the microscale characteristics are parameterized through the inclusion volume fraction $\phi$. In addition, for the fiber-reinforced RVE, the in-plane fiber orientation $\theta$ is treated as an additional design variable and optimized simultaneously. The following subsections present the corresponding results for each RVE configuration.

\subsubsection{Fiber-reinforced RVE}\label{sec:fiberRVE}

\begin{figure}[htbp]
    \centering
    \includegraphics[width=1.0 \linewidth]{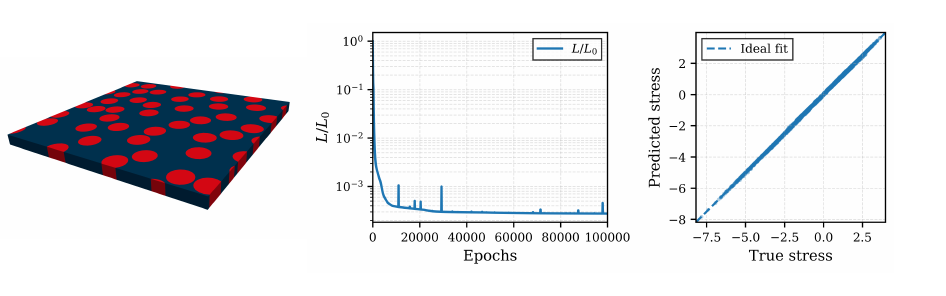}
    \caption{Fiber-reinforced representative volume element (RVE) exhibiting transverse isotropy (left), training loss evolution demonstrating convergence of the PANN model (center), and parity plot showing agreement between predicted and reference homogenized stresses (right).}
    \label{fig:fiberRVE_TrainingResults}
\end{figure}

First, we consider the RVE shown in Fig.~\ref{fig:fiberRVE_TrainingResults}. An composite reinforced with unidirectional fibers has been shown to be transversely isotropic \cite{jadoon2025inverse, Kalina2025-lq}. While a fourth-order structural tensor can be used to model the behavior of transversely isotropic materials, similar to the case of cubic anisotropy, a simpler but sufficient way to represent transverse isotropy is through a second-order structural tensor $\mathbb{N}$ which comes from Eq. \eqref{eq:TI_ST}. Thus, we aim to find the rotation $\Rb$ of a reference (canonical) direction $\eb$ to learn the preferred direction of anisotropy for our RVE.

To train the PANN surrogate, we generate a dataset containing tuples $\{\phi_i,\Cb_k, \hat{\Sb_k} \}$ with the volume fraction $\phi$ ranging from 0\% to 50\% i.e. $\phi_{\text{max}}=0.5$. Using the parameterization of the rotation tensor in Eq. \eqref{eq:rodrigues}, we find the preferred direction of anisotropy. The network learns during training that the preferred direction of anisotropy is aligned with $\nb=\{0,0,1\}$ which is consistent with the fiber orientation in Fig.~\ref{fig:fiberRVE_TrainingResults}. The evolution of the training loss and the parity plot are also shown which indicate that the network is able to accurately learn the homogenized microscopic response.

We employ this surrogate for two multiscale TO problems as explained in the following sections. The design variables considered in TO for this RVE are the macroscale material density $\bm{\rho}$, the fiber volume fraction $\bm{\phi}$, and the preferred fiber direction $\bm{\theta}$. The optimization is carried out under the constraints laid out in Eq. \eqref{eq:constraints} where we choose $\eta=0.5$ and $c=4$ in Eq. \eqref{eq:targetMass}. For both the examples presented here, we choose $\Delta p = 1$ for the continuation steps. The filter radii are chosen as $r_\mathrm{min} = 3$ for filtering the macroscale pseudo-densities, $r_\mathrm{min} = 6$ for filtering the microscale volume fractions and $r_\mathrm{min} = 2$ for filtering the fiber orientations in both examples considered in this section.

\paragraph{Simply Supported Beam}

\begin{figure}[htbp]
  \centering

    \centering
    \includegraphics[width=\linewidth]{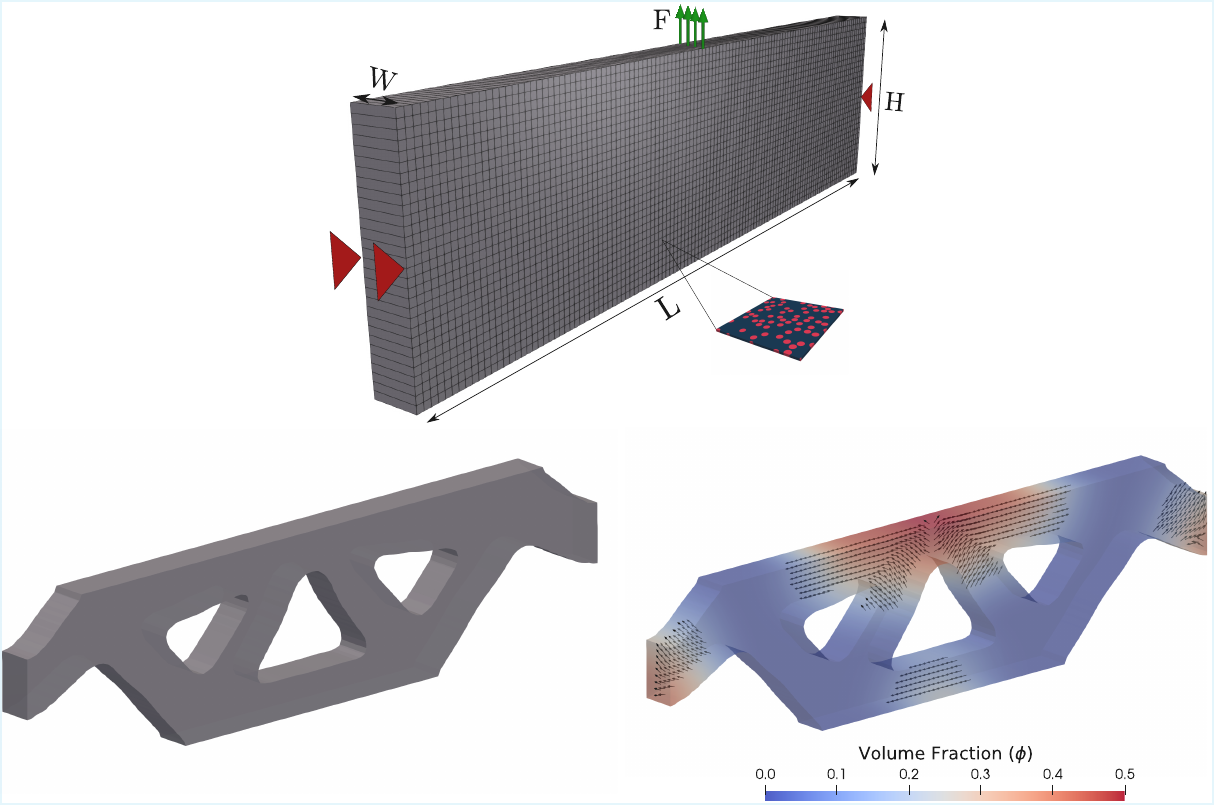}
    \label{fig:simplebeam_results}
  \caption{Topology Optimization results for the anisotropic, simply-supported beam. The preferred fiber directions in the optimized topology are shown with black arrows in the anisotropic regions.}
  \label{fig:simple_beam_results} 
\end{figure}

We implement the concurrent optimization framework on a simply supported beam subjected to a vertical load at its midspan as shown in Fig.~\ref{fig:simple_beam_results}. The beam has dimensions $L = 60\mathrm{m}$, $H = 15 \mathrm{m}$, and $W = 3\mathrm{m}$. The beam is discretized with a $50\times3\times15$ finite element mesh and is loaded with a concentrated load of $F = 500 \mathrm{N}$ applied at the top surface at the mid-point. 

\begin{figure}
    \centering
    \includegraphics[width=0.9\linewidth]{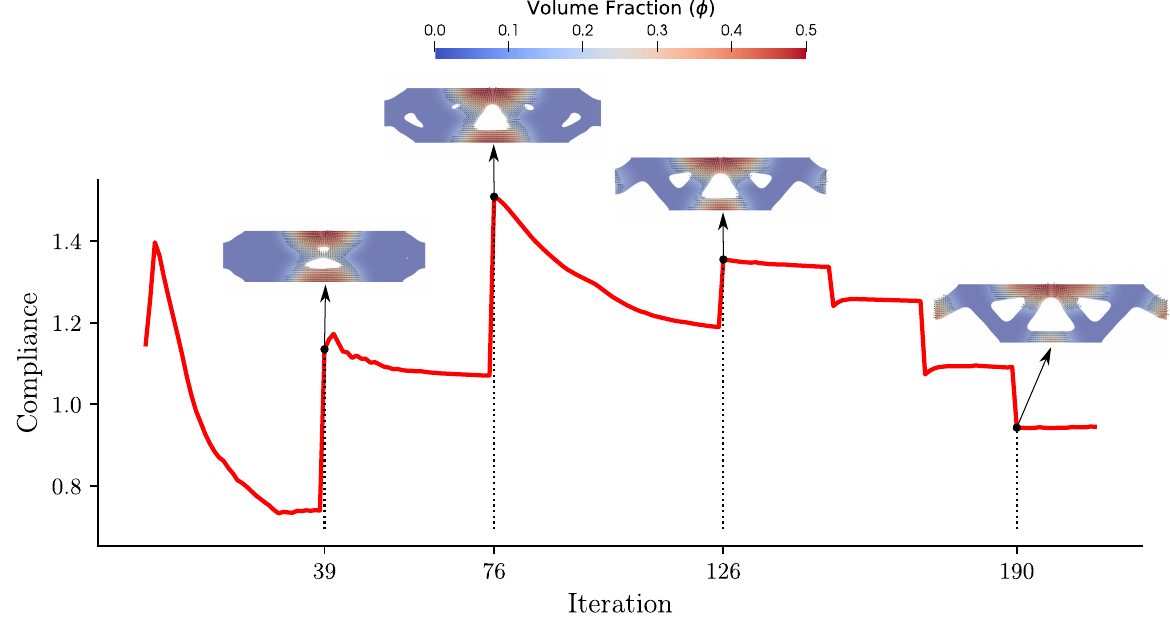}
    \caption{Compliance convergence for the simply supported beam with topology snapshots shown at iterations 39, 76, 126, and 190.}
    \label{fig:simplebeam_evolution}
\end{figure}

Fig.~\ref{fig:simple_beam_results} presents the optimization results showing the final material distribution at the macroscale and the distribution of volume fraction of fibers in this optimized topology along with the optimized orientation of the fibers. The resulting topology exhibits a truss-like load path indicating efficient material redistribution aligned with the principal stress directions. The fiber orientations are also consistent with these load paths, confirming that the framework is able to simultaneously optimize material distribution, RVE volume fraction and fiber directions. The convergence behavior during the optimization loop is shown in Fig~\ref{fig:simplebeam_evolution} along with the evolution of the topology. The sharp jumps in compliance throughout the optimization correspond to the continuation updates on the SIMP penalty exponent $p$ and the projection sharpness parameter $\beta_{\rho}$. It is observed that the optimization eliminates inefficient regions in the early iterations and by iteration 190, the design converges to a well-defined truss-like structure. 

To further assess the effect of optimizing fiber orientations, Appendix~\ref{app:fiber_direction} shows a comparison between concurrent optimization of material density and fiber orientations (Fig.~\ref{fig:rhotheta}) and fixed fiber orientations (Fig.~\ref{fig:fixedtheta}). 

\begin{figure}[htbp]
  \centering

    \centering
    \includegraphics[width=\linewidth]{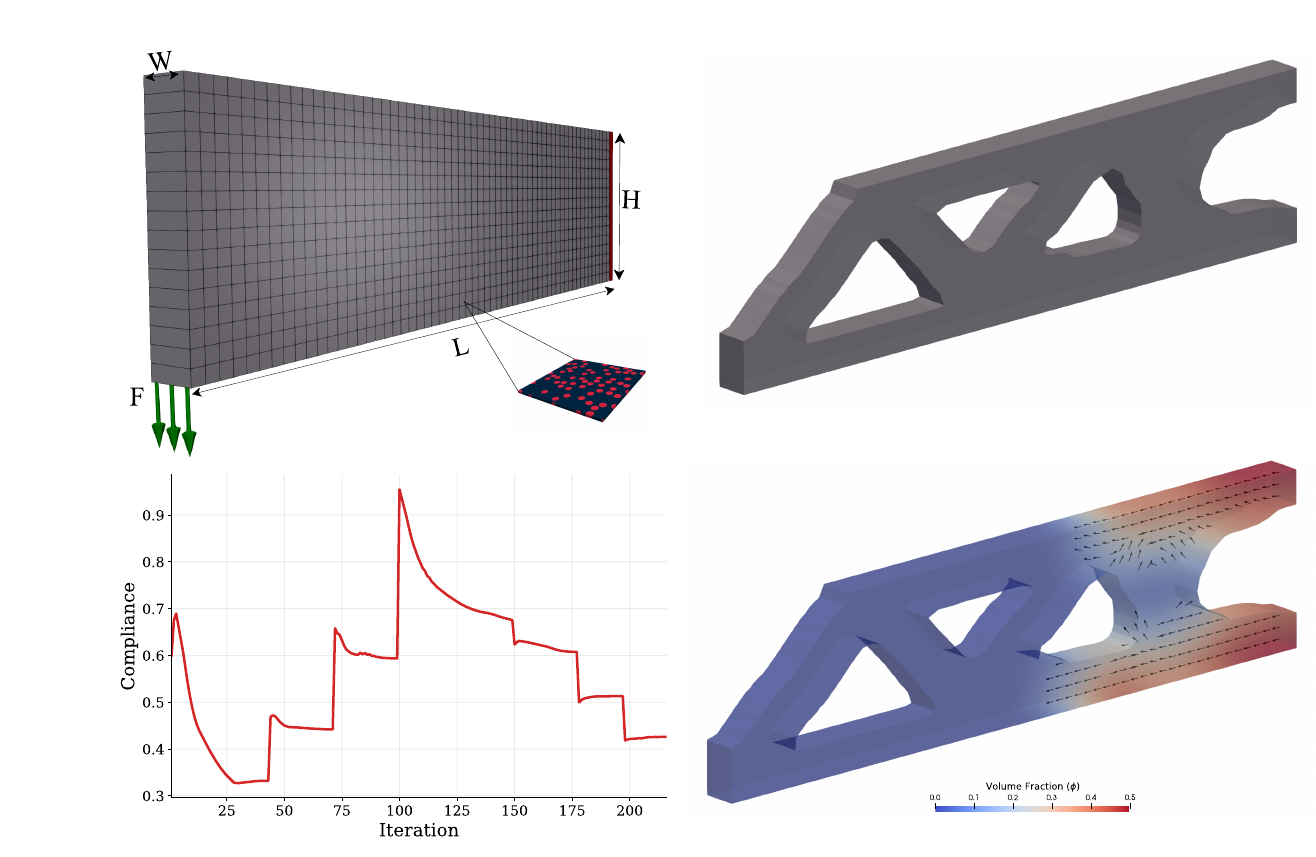}
  \caption{Topology Optimization results for the anisotropic cantilever beam.}
  \label{fig:cantilever_results} 
\end{figure}

\paragraph{Cantilever Beam}
The cantilever beam shown in Fig.~\ref{fig:cantilever_results} is subjected to a load of $F = 200$ N at the bottom of its free end. The dimensions of the beam are $L = 60$ m, $H = 15$ m, and $W = 3$ m. The optimized cantilever beam shows, as expected, a bending-dominated structural response. It is also observed that the anisotropy is strongly localized near the fixed end, where the bending stresses are highest. In these regions, the fibers are aligned with the principal stress trajectories and form continuous bands that reinforce the load path. In contrast, regions close to the free end exhibit no anisotropy as the stress state is less demanding and does not require directional reinforcement. Again, the compliance curve exhibits sharp changes , as expected, due to the continuation scheme corresponding to the updates on $p$ and $\beta_\rho$. 

To establish confidence in the ability of the proposed approach to handle geometrical nonlinearities, Appendix~\ref{app:cantilever_NL} considers the same cantilever beam under increasing load magnitudes. It is observed that as the load increases, the optimal topologies evolve significantly, reflecting the growing importance of large deformation effects. 

\subsubsection{RVE with a single spherical inclusion}\label{sec:cubicRVE}

\begin{figure}[htbp]
    \centering
    \includegraphics[width=1.0 \linewidth]{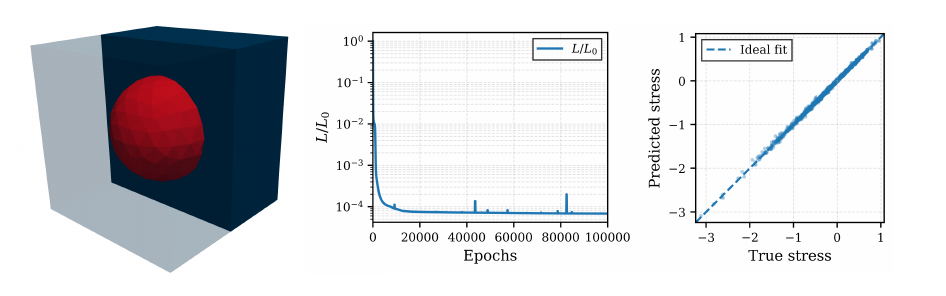}
    \caption{Representative volume element (RVE) with a spherical inclusion exhibiting cubic anisotropy (left), training loss evolution demonstrating convergence of the PANN model (center), and parity plot showing agreement between predicted and reference homogenized stresses (right).}
    \label{fig:cubicRVE_TrainingResults}
\end{figure}

Next, we consider an RVE with a single spherical inclusion, shown in Fig.~\ref{fig:cubicRVE_TrainingResults}, with the ratio of shear modulii of the inclusion to the surrounding matrix material equal to $10^3$. Such an RVE has been shown \cite{Kalina2025-lq} to exhibit cubic anisotropy whereas for a much smaller ratio of shear modulii, it exhibits isotropic response \cite{jadoon2025inverse}. We require a fourth-order structural tensor to be able to model the mechanical response of this RVE. Several other anisotropy groups can also be represented by fourth-order structural tensors \cite{Kalina2025-lq} and hence this example aims to prove the applicability of our framework for all such anisotropy classes.

\begin{figure}[htbp]
    \centering
    \includegraphics[width=1.0 \linewidth]{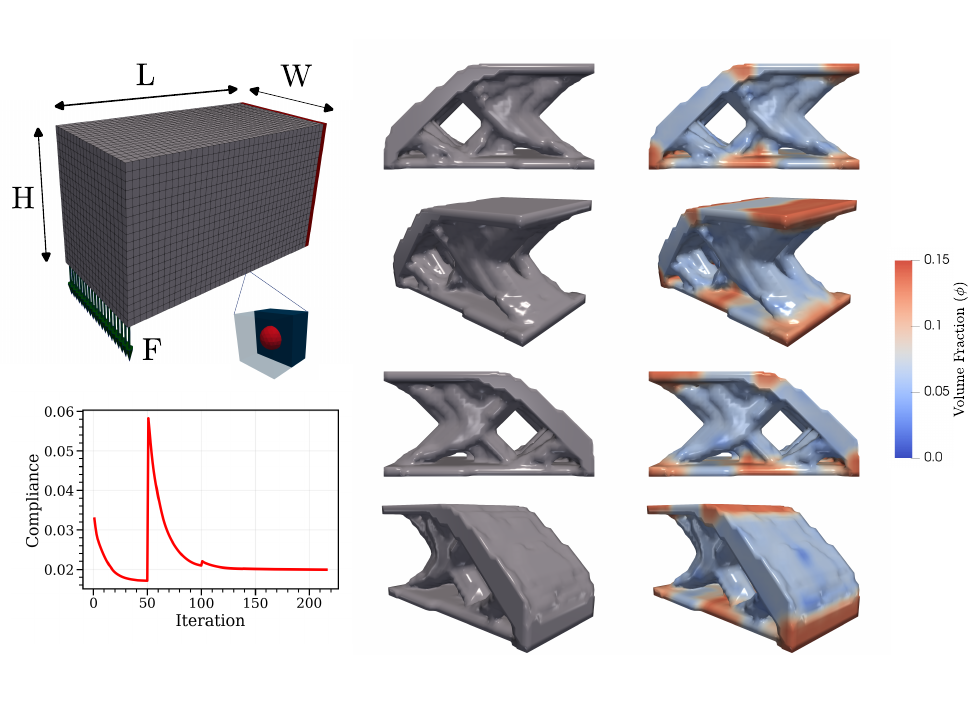}
    \caption{Topology optimization results for the compressible case with a cubic-anisotropic RVE. The problem setup, including the design domain, boundary conditions, and associated RVE, is shown together with the compliance evolution over the optimization process. The final optimized density distribution is presented along with the corresponding spatial distribution of the microstructural volume fraction $\phi$. Multiple views of the optimized design are provided to enhance visualization.}
    \label{fig:cubicRVE_TopOptResults}
\end{figure}

In order to obtain a surrogate representation for the homogenized response of this RVE, we first generate a dataset consisting of tuples $\{\phi_i,\Cb_k, \hat{\Sb_k} \}$ where we choose $\phi_{\text{max}}=0.25$ which is controlled through the radius of the spherical inclusion. Here, $k=1,\ldots,N_F$ where $N_F$ represents the number of deformation gradients sampled. We choose $N_F=200$ and sample the deformation gradients via Latin Hypercube Sampling (LHS) \cite{stein1987large} within a range corresponding to up to 20\% strain. To obtain the stresses $\hat\Sb_k$ corresponding to each deformation gradient $\Fb_k$, the prescribed deformation was applied to the microstructural RVE under periodic boundary conditions, and the resulting stress field was homogenized to compute $\hat\Sb_k$. Additionally, we parametrize the rotation tensor $\Rb$ using Eq. \eqref{eq:rodrigues} and learn the orientation of the structural tensor $\mathbb{A}$ (see Eq. \eqref{eq:cubic_ST}). The evolution of loss during training along with the parity plot at the end of training is shown in Fig.~\ref{fig:cubicRVE_TrainingResults}. We can see a close match between the predicted and true stresses. Additionally, the network learns during training $\Rb=\Ib$, implying $\mathbb{A}$ is aligned with the canonical bases, i.e., $\mathbb{A} = \sum \mathbf{e}_i \otimes\mathbf{e}_i \otimes\mathbf{e}_i \otimes\mathbf{e}_i$.

Consider the 3D cantilever beam shown in Fig.~\ref{fig:cubicRVE_TopOptResults} embedded with the cubic RVE under consideration. The dimensions of the beam are $L = 20$ m, $H = 10$ m, and $W = 10$m and is discretized with a $40\times20\times20$ mesh. The bottom of the beam's free end is loaded with a distributed force of $F=4$N/m. Using the trained ISNNs to give the homogenized microscopic response, we solve the macroscale BVP and perform concurrent optimization of the macroscale topology with $\eta=0.2$ in Eq. \eqref{eq:constraints}, the macroscale densities $\rho_e$ bounded between 0 and 1, and the volume fraction of inclusions bounded between 0\% and 25\%. The target mass in Eq. \eqref{eq:targetMass} is chosen with $c=3$. Additionally, $r_{\mathrm{min}} = 0.5$ in Eq. \eqref{eq:Rmin} for filtering the macroscale pseudo-densities whereas $r_{\mathrm{min}} = 1.0$ to filter the microscale volume fractions.

The optimization results are shown in Fig.~\ref{fig:cubicRVE_TopOptResults}. The optimized topology exhibits a clear load-bearing path connecting the loaded and fixed boundaries, with material concentrated along regions of high stress, as expected for compliance minimization. In addition to the macroscale density distribution, the optimization simultaneously adapts the microstructural volume fraction $\phi$, while the orientation of the anisotropy is fixed in this example, resulting in a spatially varying anisotropic response across the domain. It can be observed that higher values of $\phi$ are predominantly localized near the boundaries and regions experiencing higher stresses, indicating that the optimizer leverages the increased stiffness contrast provided by the inclusion phase in critical regions. Conversely, regions with lower stress levels exhibit reduced inclusion volume fractions, guided by the constraint on the inclusion mass. The compliance evolution demonstrates stable convergence under the proposed continuation scheme, with a rapid initial decrease followed by gradual refinement of the design. The sharp jumps in compliance during optimization correspond to the continuation updates on the SIMP penalty exponent $p$ and the projection sharpness parameter $\beta_{\rho}$. Overall, these results highlight the ability of the proposed framework to effectively perform concurrent optimization across scales, tailoring both the macroscale topology and the underlying microstructural volume fraction to achieve improved structural performance.

Another set of results is presented in Appendix \ref{app:cubicRVE_fixedPhiComparison}, where we compare the proposed concurrent multiscale optimization approach with cases involving prescribed microstructural configurations. Specifically, we consider an isotropic case with $\phi = 0$ and an anisotropic case where $\phi \neq 0$ is enforced to be spatially uniform across all elements. In the latter, the value of $\phi$ is chosen such that the total mass of inclusions matches the target mass considered here. These results are compared against the fully coupled case presented here, where both the macroscale topology and the microscale inclusion volume fraction are optimized simultaneously. The results indicate almost $8\%$ reduction in end compliance when optimizing both the macroscale topology and the microscale distribution of inclusion volume fraction.

\subsection{Nearly incompressible case}\label{sec:incompRVE}

\begin{figure}[htbp]
    \centering
    \includegraphics[width=1.0 \linewidth]{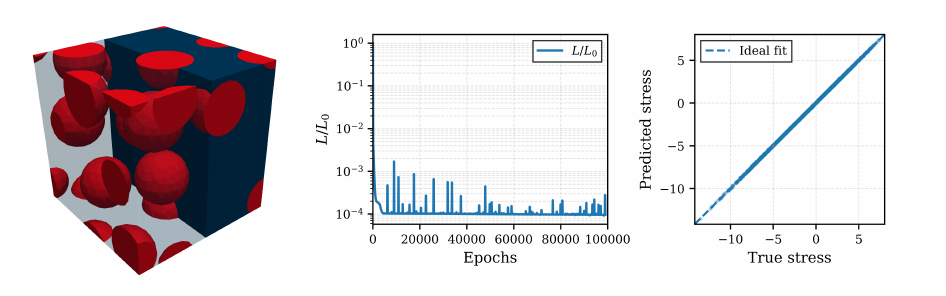}
    \caption{Representative volume element (RVE) with randomly distributed inclusions exhibiting isotropic and nearly incompressible behavior (left), training loss evolution demonstrating convergence of the PANN model (center), and parity plot showing agreement between predicted and reference homogenized stresses (right).}
    \label{fig:incompRVE_TrainingResults}
\end{figure}

For the case of nearly incompressible hyperelasticity, we consider the RVE presented in Fig.~\ref{fig:incompRVE_TrainingResults} whose homogenized response is assumed to be isotropic. Consequently, no structural tensors are required, and the constitutive behavior is fully characterized by invariants of the isochoric right Cauchy-Green tensor $\bar{\Cb}$. In contrast to the compressible anisotropic case, we therefore restrict the surrogate model to represent only the isochoric part of the free energy, while the volumetric response is treated analytically through the pressure field as described in Section \ref{Sec:Theory}. This surrogate modeling approach has been validated in Appendix~\ref{App:PhenvsPANNvalidation}, where a PANN-based constitutive model is compared against the Mooney-Rivlin model. In the present example, only the macroscale density field is treated as a design variable, and no microstructural parameters are optimized. Nevertheless, the proposed framework readily extends to parametrized microstructures in the nearly incompressible setting, following the same approach as in the compressible case. In particular, quantities such as the inclusion volume fraction can be introduced as additional design variables and optimized concurrently, enabling fully coupled multiscale topology optimization.

Following Ref. \cite{spencer2026monolithiccomputationalhomogenizationframework}, nearly incompressible homogenization was modeled using a $\bar{J}$-averaging technique, whereby the volumetric response is controlled through an averaged Jacobian measure. This treatment enforces near-incompressibility while avoiding the volumetric locking that commonly arises in standard displacement-based formulations. Using this scheme, a dataset of tuples $\{\bar{\mathbf{C}}_k, \hat{\mathbf{S}}_k\}$ was generated through computational homogenization.

\begin{figure}[htbp]
    \centering
    \includegraphics[width=1.0 \linewidth]{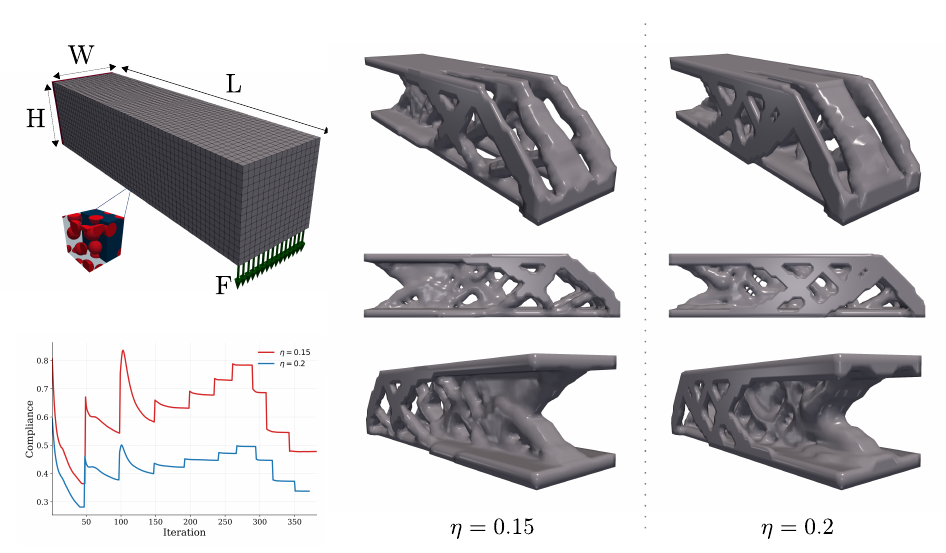}
    \caption{Topology optimization results for the nearly incompressible isotropic case. The problem setup and compliance evolution are shown for two values of the volume constraint parameter $\eta$. The corresponding optimized density distributions for $\eta = 0.15$ and $\eta = 0.2$ are presented, highlighting the influence of the available material volume on the resulting structural topology.}
    \label{fig:incompRVE_TopOptResults}
\end{figure}

ISNNs were then subsequently trained on this dataset using the framework presented in Section \ref{Sec:Model}. Fig.~\ref{fig:incompRVE_TrainingResults} shows the evolution of training losses over epochs as well as the parity plot between the predicted and true stresses. These results indicate that ISNN-based PANNs are able to accurately represent the homogenized isochoric free energy. We embed this trained surrogate into the cantilever beam shown in Fig.~ \ref{fig:incompRVE_TopOptResults} which has dimensions $L = 40$ m, $H = 10$ m, and $W = 10$m and is discretized with a $60\times15\times15$ mesh. The bottom of the beam's free end is loaded with a distributed force of $F=100$ N/m. For this problem, we set $\Delta p =0.5$ for continuation strategy and $r_{\mathrm{min}} = 0.75$ in Eq. \eqref{eq:Rmin} for filtering the density. We then solve the topology optimization problem using the mixed formulation as explained in Section \ref{Sec:TO_problem}. Here, $\kappa_0$ is chosen based on a reference nearly incompressible linear elastic material, computed as $\kappa_0 = E / \bigl(3(1 - 2\nu)\bigr)$ with $E = 700\,\text{MPa}$ and $\nu = 0.499$, ensuring a sufficiently stiff volumetric response to enforce near-incompressibility. The Young’s modulus $E$ is selected to provide a stiffness scale consistent with the homogenized response. We optimize the topologies for two different target macroscale volume fractions. Specifically, we solve for $\eta=0.15$ and $\eta=0.2$ in Eq. \eqref{eq:constraints}. Since we only optimize the macroscale topology, the second constraint $g_2$ in Eq. \eqref{eq:constraints} is inactive for this case. The optimized topologies along with the compliance evolution for both these cases are shown in Fig.~ \ref{fig:incompRVE_TopOptResults}.

We can see that compliance evolution shows a similar trend as the examples presented in previous sections. Specifically, we see sharp jumps in compliance during the optimization process which correspond to the continuation updates on the SIMP penalty exponent $p$ and the projection sharpness parameter $\beta_{\rho}$. The results clearly demonstrate the influence of the prescribed macroscale volume fraction on the optimized structural response. As expected, the design corresponding to $\eta = 0.2$ exhibits a lower compliance compared to the $\eta = 0.15$ case, owing to the increased availability of material that facilitates the formation of stronger and more continuous load-bearing paths. In contrast, the stricter volume constraint leads to a more sparse and slender topology, which reduces structural stiffness and increases compliance. These observations are consistent with the compliance evolution shown in Fig.~\ref{fig:incompRVE_TopOptResults}, and highlight the ability of the proposed framework to adapt the topology in a physically meaningful manner under different material constraints.

\section{Discussion and conclusions}
\label{Sec:discussion}

In this work, we have presented a concurrent multiscale topology optimization framework for nonlinear hyperelastic materials, leveraging physics-augmented neural networks (PANNs) as surrogate constitutive models. The proposed approach enables simultaneous optimization of macroscale material distribution and microscale features, while maintaining thermodynamic constraints, mechanistic assumptions, and numerical stability. The numerical examples demonstrate the versatility of the framework across different material classes and constitutive settings. For the compressible anisotropic cases, both the fiber-reinforced (transversely isotropic) and single spherical inclusion (cubic anisotropic) RVEs exhibit clear coupling between macroscale topology and microscale design variables. In particular, the optimization consistently exploits microstructural parameters such as the inclusion volume fraction or fiber orientation to enhance stiffness in critical regions. This highlights the effectiveness of the framework in distributing anisotropies spatially in a physically meaningful manner, rather than prescribing it a priori.

For the fiber-reinforced case, the additional degree of freedom associated with the orientation field enables the optimizer to align the preferred direction of anisotropy with dominant stress paths, leading to improved structural performance. Similarly, for the cubic anisotropic RVE, the spatial variation of the inclusion volume fraction reflects the optimizer’s ability to selectively stiffen regions of high load demand while maintaining material efficiency elsewhere. These observations confirm that the proposed framework successfully captures the interplay between macroscale structural behavior and microscale material design. In contrast, the nearly incompressible isotropic case demonstrates the applicability of the framework to materials with strong volumetric constraints. By decoupling the volumetric and isochoric responses and employing a mixed formulation, the method avoids volumetric locking while maintaining stable and accurate solutions. The resulting topologies exhibit expected behavior under different volume fraction constraints, with increased material availability leading to more robust load-bearing structures and reduced compliance. These results further validate the flexibility of the proposed approach in handling fundamentally different constitutive regimes within a unified framework.

A key advantage of the proposed method lies in the use of PANNs to replace the microscale boundary value problem. By embedding physical constraints directly into the neural network architecture, the resulting surrogate models remain thermodynamically admissible and numerically stable. This is particularly important in nonlinear settings, where conventional black-box models may lead to unphysical responses or convergence issues. At the same time, the use of PANNs significantly reduces the computational cost compared to classical FE\textsuperscript{2} approaches, making multiscale topology optimization tractable for large-scale nonlinear problems.

Despite these advantages, the present framework also has several limitations that suggest directions for future work. First, the surrogate models are trained offline on RVEs assuming periodic boundary conditions and a clear separation of scales. In settings where scale separation is weak or microstructures are non-periodic, additional modeling effort may be needed. Second, the current framework assumes prior knowledge of the material symmetry class, which may not always be available in practice. Extending the approach to automatically infer symmetry classes or handle more complex microstructures represents an important direction for future research. Third, the current work is restricted to rate-independent, hyperelastic materials. However, this is only a modeling choice and extensive research is being done to develop PANN-based frameworks to model inelasticity or multi-physics behavior (see e.g.  Refs.\cite{rosenkranz2024viscoelasticty, jadoon2025automated, kalina_viscoelasticity, fuhg2024polyconvex, klein2022finite, Reese_TVP}) which can be used to extend the proposed framework to incorporate complex material behavior. Finally, we do not explicitly incorporate manufacturing constraints at the microscale. Embedding additional constraints or regularization terms that encode process-specific requirements would be necessary for direct deployment in additive manufacturing or fiber-placement technologies. In the same spirit, uncertainty in material properties, microstructural geometry, or loading has not been addressed here but could be handled by combining the present framework with robust or reliability-based topology optimization and uncertainty quantification approaches \cite{dirt, feng2025finite, robust_2}. 

Overall, the proposed concurrent optimization framework demonstrates that parametrized physics-augmented neural networks, and ISNNs in particular, provide a viable and flexible bridge between microscale anisotropic homogenization and macroscale topology optimization in the finite strain regime. By encoding physical constraints at the architecture level and exploiting invariant-based representations, these models deliver stable constitutive behavior and accurate sensitivities, enabling non-trivial improvements in structural performance through simultaneous design of topology and anisotropic microstructure. We anticipate that extending this approach to more general material behaviors, multiple microstructure families, and manufacturing-aware constraints will further enhance its potential for the practical design of architected and anisotropic materials.

\section*{Acknowledgments}
This work was funded by the Laboratory Directed Research and Development (LDRD) program at Sandia National Laboratories; this funding is gratefully acknowledged.
Sandia National Laboratories is a multi-mission laboratory managed and operated by National Technology \& Engineering Solutions of Sandia, LLC (NTESS), a wholly owned subsidiary of Honeywell International Inc., for the U.S. Department of Energy’s National Nuclear Security Administration (DOE/NNSA) under contract DE-NA0003525. This written work is authored by an employee of NTESS. The employee, not NTESS, owns the right, title and interest in and to the written work and is responsible for its contents. Any subjective views or opinions that might be expressed in the written work do not necessarily represent the views of the U.S. Government. The publisher acknowledges that the U.S. Government retains a non-exclusive, paid-up, irrevocable, world-wide license to publish or reproduce the published form of this written work or allow others to do so, for U.S. Government purposes. The DOE will provide public access to results of federally sponsored research in accordance with the DOE Public Access Plan. \\
This material is based upon work partially supported by the U.S. National Science Foundation under award No. 2452029 and partially under award No. 2235856 (MKR).  
The opinions, findings, and conclusions, or recommendations expressed are those of the authors and do not necessarily reflect the views of the NSF.
\\
The authors acknowledge the Texas Advanced Computing Center (TACC) at The University of Texas at Austin for providing computational resources that have contributed to the research results reported within this paper. 

\section*{Data availability}
The code will be made available after the acceptance of this manuscript.

\clearpage
\appendix
\numberwithin{equation}{section}
\numberwithin{figure}{section}

\textbf{\large Appendix}

\section{Verification of adjoint sensitivities using central finite differences}
\label{app:adjoint_verification}
To verify the correctness of the adjoint-based sensitivity implementation, we compare adjoint directional derivatives presented in Section \ref{sec:sensitivity} against finite-difference approximations computed along randomly generated perturbation directions. The verification is performed for both the compressible and incompressible formulations. In the compressible case, the domain consists of a cantilever beam of dimensions $L=10$, $W=3$, and $H=3$, discretized using a $10 \times 1 \times 3$ mesh. The fiber-reinforced RVE is considered, where the macroscale pseudo-density and the microscale parameters governing the anisotropy are treated as design variables. The corresponding set of design variables is given by:

\begin{equation}
   \bm{\Theta}_{\mathrm{comp}} = \{\bm{\rho}, \bm{\phi}, \bm{\theta}\}. 
\end{equation}

In contrast, for the incompressible formulation, only the macroscale pseudo-density is optimized, and the design variable set reduces to:

\begin{equation}
    \bm{\Theta}_{\mathrm{inc}} = \{\bm{\rho}\}.
\end{equation}

Additionally, for the nearly incompressible case, a cantilever beam with dimensions $L=20$, $W=3$, and $H=5$ is considered, discretized using a $20 \times 1 \times 5$ mesh. Let the compliance be denoted by $C(\bm{\Theta})$. The adjoint formulation provides the gradient $\partial C / \partial \bm{\Theta}$, which is verified here through directional derivatives. For each test, we construct an element-wise random perturbation vector $\db \in \mathbb{R}^{N_{\mathrm{elem}}}$ drawn from a standard normal distribution. This vector is normalized according to:

\begin{equation}
    \hat{\db} = \frac{\db}{\max |\db|},
\end{equation}

such that its largest entry has unit magnitude. This normalization ensures that the perturbation remains uniformly bounded across all elements. The normalized direction is then scaled to ensure that the perturbed design variables remain within their admissible bounds. For the compressible case, the resulting perturbation vector can be written as:

\begin{equation}
    \db^\star = \{\gamma_{\rho} \hat{\db}, \gamma_{\phi} \hat{\db}, \gamma_{\theta} \hat{\db} \}^{T},
\end{equation}

where only the scaling coefficient corresponding to the active design variable is taken to be nonzero. For the incompressible case, the perturbation reduces to:

\begin{equation}
    \db^\star = \{\gamma_{\rho} \hat{\db}\}^{T}.
\end{equation}

For a given perturbation direction $\db^\star$, the directional derivative predicted by the adjoint sensitivities is given by:

\begin{equation}
g_{\mathrm{adj}} =
\frac{\partial C}{\partial \bm{\Theta}} \cdot \db^\star.
\label{eq:dir_adj}
\end{equation}

As a reference, we compute the directional derivative using central finite differences:

\begin{equation}
g_{\mathrm{FD}}(h) =
\frac{
C(\bm{\Theta} + h \db^\star)
-
C(\bm{\Theta} - h \db^\star)
}{2h},
\label{eq:central_fd_appendix}
\end{equation}
where $h$ denotes the finite-difference step size. Central differences are employed instead of forward differences due to their second-order accuracy, which reduces truncation error and improves robustness when verifying sensitivities in nonlinear finite-strain settings. The accuracy of finite differences depends strongly on the choice of step size $h$. To assess the robustness of the adjoint sensitivities, we evaluate the directional derivative over a range of step sizes spanning several orders of magnitude. Specifically, we consider 15 logarithmically spaced values between $10^{0}$ and $10^{-8}$. Furthermore, the verification is performed over $N_d = 15$ independent random perturbation directions. For each step size $h$, the discrepancy between the adjoint and finite-difference directional derivatives is quantified by:

\begin{equation}
\epsilon(h)
=
\left| g_{\mathrm{FD}}(h) - g_{\mathrm{adj}} \right| \ .
\label{eq:error_metric_appendix}
\end{equation}

\begin{figure}[htbp]
    \centering
    \begin{subfigure}[t]{0.45\linewidth}
        \centering
        \includegraphics[width=\linewidth]{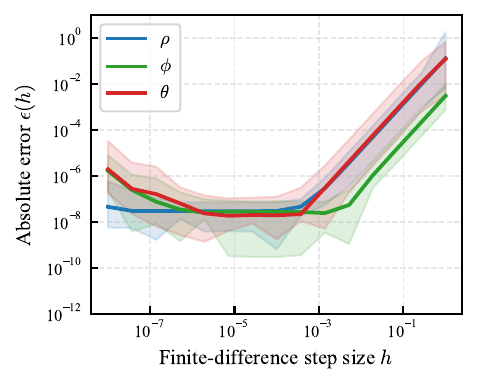}
        \caption{}
        \label{fig:FDvsAdjoint_compressible}
    \end{subfigure}
    \hfill
    \begin{subfigure}[t]{0.45\linewidth}
        \centering
        \includegraphics[width=\linewidth]{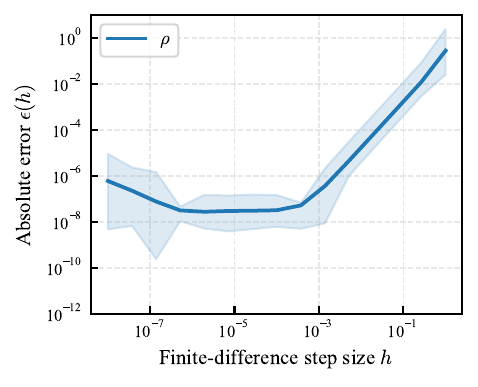}
        \caption{}
        \label{fig:FDvsAdjoint_incompressible}
    \end{subfigure}
    \caption{Verification of adjoint sensitivities using central finite differences. In the (a) compressible case, the design variables are $\bm{\rho}$, $\bm{\phi}$, and $\bm{\theta}$, whereas in the (b) incompressible case only $\bm{\rho}$ is present. Solid lines denote the median error over $N_d=15$ random perturbation directions, and the shaded regions indicate the corresponding minimum--maximum range.}
    \label{fig:FDvsAdjoint_subfigures}
\end{figure}

Fig.~\ref{fig:FDvsAdjoint_subfigures} summarizes the verification results for both formulations. Fig.~\ref{fig:FDvsAdjoint_compressible} corresponds to the compressible case and reports the absolute error $\epsilon(h)$ for the three design variable fields $\bm{\rho}$, $\bm{\phi}$, and $\bm{\theta}$. Fig.~\ref{fig:FDvsAdjoint_incompressible} shows the corresponding result for the incompressible case, where only $\bm{\rho}$ is present. In both subfigures, the solid lines denote the median error over the $N_d=15$ random perturbation directions, while the shaded regions indicate the corresponding minimum-maximum range. For all considered design variables, the error exhibits the expected U-shaped behavior. For large step sizes, the error is dominated by truncation effects and decreases as $h$ is reduced. An optimal range of step sizes is then reached, typically between $h \sim 10^{-3}$ and $10^{-5}$, where the error attains its minimum and excellent agreement between adjoint and finite-difference directional derivatives is observed. This regime confirms that the adjoint sensitivities are consistent with the fully linearized nonlinear equilibrium equations. For smaller step sizes, the error increases again due to round-off effects and the finite tolerance of the nonlinear equilibrium solver.

\section{Single Scale Topology Optimization}\label{App:PhenvsPANNvalidation}
\subsection{Compressible Isotropic Hyperelastic}
\begin{figure}[htbp]
    \centering
    \includegraphics[width=1.0\linewidth]{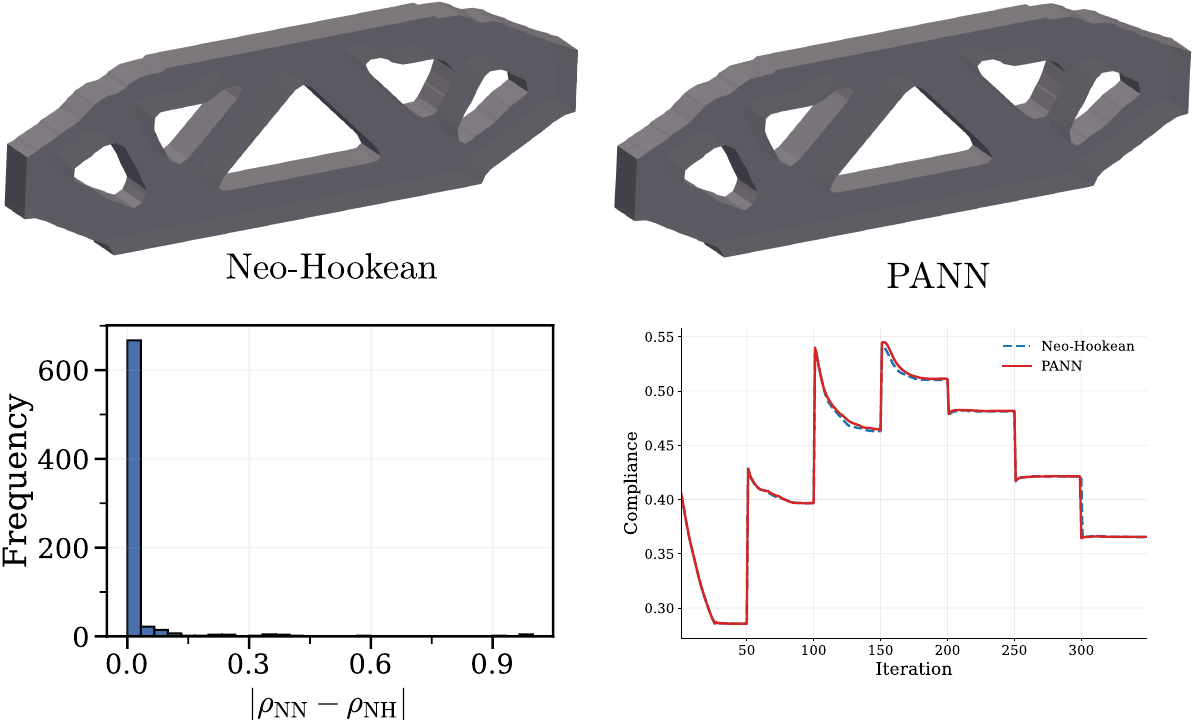}
    \caption{Comparison of convergence history and optimized structure for the single scale TO between phenomenological and ISNN-based material model for compressible isotropic hyperelastic case.}
    \label{fig:NHvsNN}
\end{figure}

In this Appendix, we validate the performance and accuracy of ISNN by comparing it to a classical phenomenological material model within the TO framework for the compressible isotropic hyperelastic case. This validation is performed to check whether the ISNN-based TO framework can replace analytical constitutive laws without introducing numerical artifacts or degrading accuracy.

For this purpose, we generate a synthetic dataset using a compressible Neo-Hookean material model. We sample $N = 500$ deformation gradients $\mathbf{F}$ using LHS\cite{stein1987large} within a range of up to $20\%$ strain. The corresponding right Cauchy–Green tensors $\mathbf{C} = \mathbf{F}^T \mathbf{F}$ are then used to evaluate the strain energy density.

To model compressible isotropic hyperelasticity, we employ the standard Neo-Hookean strain energy function:
\begin{equation}
    \psi(\mathbf{C}) = \frac{1}{2} c_1 (I_1 - 3) - c_1 \log J + \frac{1}{2} c_2 (J - 1)^2,
\end{equation}
where $I_1 = \mathrm{tr}(\mathbf{C})$ and $J = \sqrt{\det(\mathbf{C})}$. The material parameters are fixed as $c_1 = 10.0$ and $c_2 = 5.0$.

This strain energy function yields the second Piola–Kirchhoff stress tensor of the form:
\begin{equation}
    \mathbf{S} = c_1 \left( \mathbf{I} - \mathbf{C}^{-1} \right) + c_2 (J - 1)\, J \, \mathbf{C}^{-1}.
\end{equation}

We utilize the same simple beam benchmark described in \ref{sec:fiberRVE} for this comparison. As illustrated in Fig.~\ref{fig:NHvsNN}, the optimized topologies generated by both material models exhibit nearly identical structures. The load-bearing members and void distributions show close agreement, indicating that the ISNN accurately captures the stress response and associated tangent information. Furthermore, the compliance evolution in the ISNN-based TO framework remains consistent with that obtained using the phenomenological Neo-Hookean model.

\subsection{Incompressible Isotropic Hyperelastic}
\begin{figure}[htbp]
    \centering
    \includegraphics[width=1.0\linewidth]{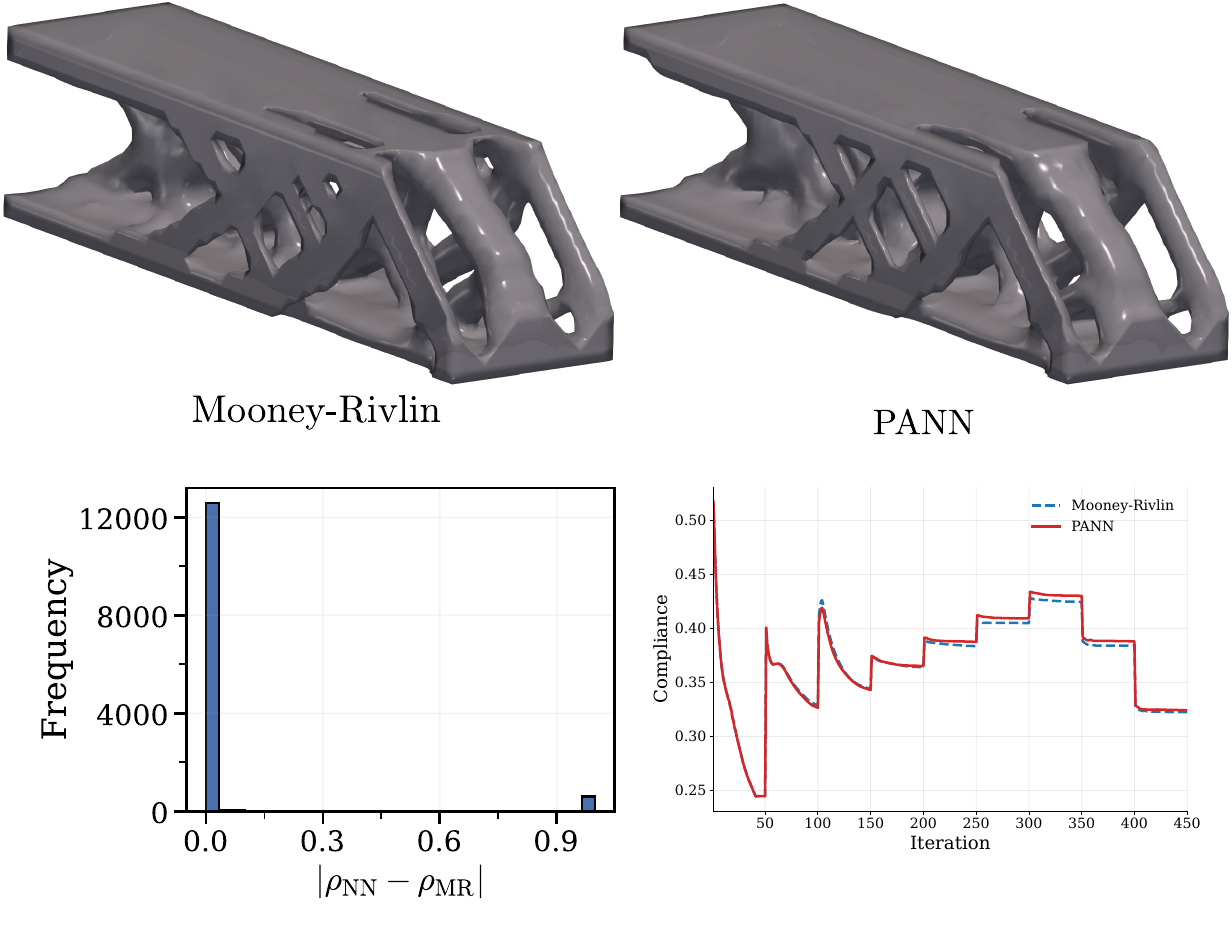}
    \caption{Comparison of convergence history and optimized structure for the single scale TO between phenomenological and ISNN-based material model for incompressible isotropic hyperelastic case.}
    \label{fig:MRvsNN}
\end{figure}

For the incompressible isotropic hyperelastic case, we generate the synthetic data using an incompressible Mooney-Rivlin model. We again sample $N = 500$ deformation gradients $\mathbf{F}$ using LHS within a $20\%$ strain range. The material parameters are derived from a nearly incompressible linear elastic response with Young’s modulus $E = 7 \times 10^2$ MPa and Poisson’s ratio $\nu = 0.499$. The corresponding shear modulus is given by:
\begin{equation}
\mu = \frac{E}{2(1+\nu)},
\end{equation}
and the Mooney--Rivlin parameters are chosen as $C_{10} = 0.25\,\mu$ and $C_{01} = 0.25\,\mu$.

To enforce incompressibility, we employ the isochoric formulation of the Mooney--Rivlin strain energy function, defined in terms of the modified invariants $\bar{I}_1$ and $\bar{I}_2$ as:
\begin{equation}
\psi(\bar{I}_1, \bar{I}_2) = C_{10}(\bar{I}_1 - 3) + C_{01}(\bar{I}_2 - 3),
\end{equation}
where the invariants are computed from the isochoric right Cauchy--Green tensor $\bar{\mathbf{C}} = J^{-2/3}\mathbf{C}$. 

Fig.~\ref{fig:MRvsNN} shows the comparison between the optimized topology and the compliance evolution between the Mooney-Rivlin and the PANN-based material models.

\section{Isotropic vs anisotropic comparison} \label{app:cubicRVE_fixedPhiComparison}
In this section, we assess the performance of the proposed framework when the microscale design is fixed, thereby reducing the problem to a classical topology optimization setting at the macroscale. The results are compared against the concurrent multiscale optimization presented in Section \ref{sec:cubicRVE}, where both the macroscale topology and the microscale inclusion volume fraction are treated as design variables. For a fair comparison, the geometry, mesh, loading conditions, as well as topology optimization hyperparameters are kept the same as those reported in Section \ref{sec:cubicRVE}.

\begin{figure}[htbp]
  \centering
  \begin{subfigure}{1.0\textwidth}
    \centering
    \includegraphics[width=\linewidth]{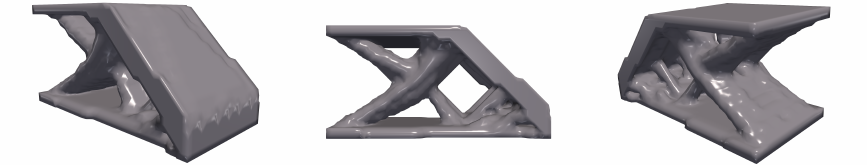}
    \caption{$\phi=0$}
    \label{fig:fixed_iso}
  \end{subfigure}\hfill
  \begin{subfigure}{1.0\textwidth}
    \centering
    \includegraphics[width=\linewidth]{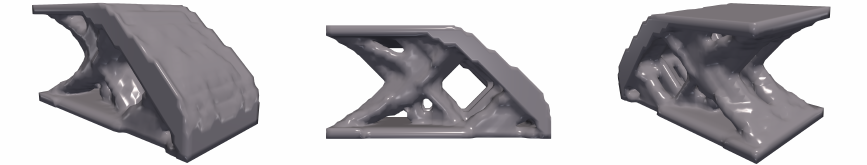}
    \caption{Fixed $\phi \neq 0$}
    \label{fig:fixed_cubic}
  \end{subfigure}

  \caption{Optimization results for the cantilever beam with fixed microstructures.}
  \label{fig:cubicRVEresults_fixed}
\end{figure}

Specifically, we consider two cases. In the first case, the RVE is isotropic with no inclusions, i.e., $\phi = 0$. In the second case, an RVE is considered with a fixed inclusion volume fraction $\phi \neq 0$, which is enforced to be spatially uniform across all elements. The value of $\phi$ is chosen such that the total mass of inclusions matches the target mass used in the concurrent optimization problem. The resulting optimized topologies for both cases are shown in Fig.~\ref{fig:cubicRVEresults_fixed}.

\begin{figure}[H]
    \centering
    \includegraphics[width=0.7\linewidth]{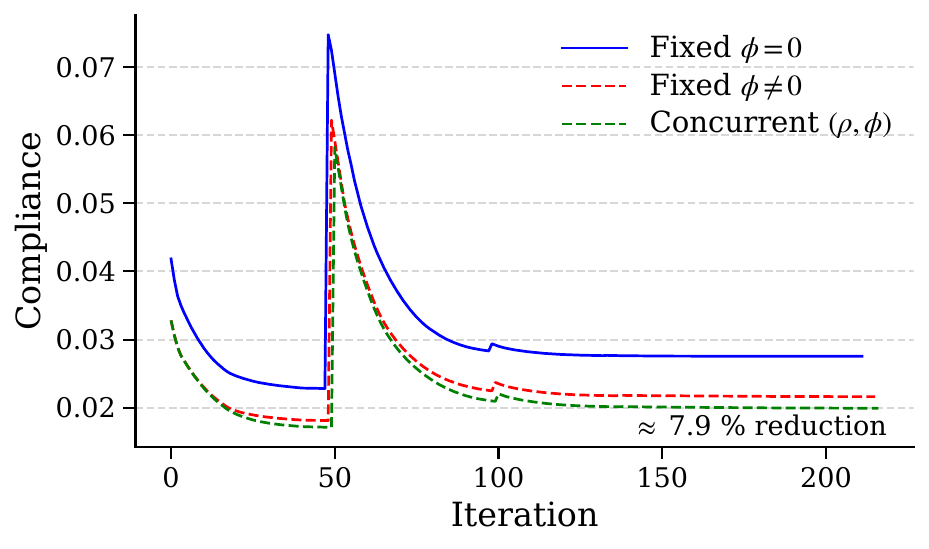}
    \caption{Comparison of compliance evolution between the anisotropic and isotropic cases.}
    \label{fig:iso_cubic_compliance}
\end{figure}

To quantify these differences, Fig.~\ref{fig:iso_cubic_compliance} compares the compliance evolution for the isotropic case, the fixed anisotropic case, and the fully coupled concurrent optimization. The anisotropic case with fixed $\phi$ consistently outperforms the isotropic case due to the enhanced stiffness introduced by the inclusions. More importantly, the concurrent optimization of both $\rho$ and $\phi$ yields the lowest compliance among all cases, achieving approximately $8\%$ reduction compared to the fixed anisotropic configuration.

These results highlight the importance of concurrent multiscale optimization. While introducing anisotropy through a fixed microstructure improves performance relative to the isotropic case, the ability to spatially vary the inclusion volume fraction provides an additional degree of freedom that enables the optimizer to allocate stiffness more effectively in critical regions. This leads to improved structural performance without increasing the total material usage, demonstrating the advantage of the proposed framework over traditional approaches with prescribed microstructures.

\section{Effect of Optimizing Fiber Direction}
\label{app:fiber_direction}

\begin{figure}[H]

    \centering
    \includegraphics[width=0.75\linewidth]{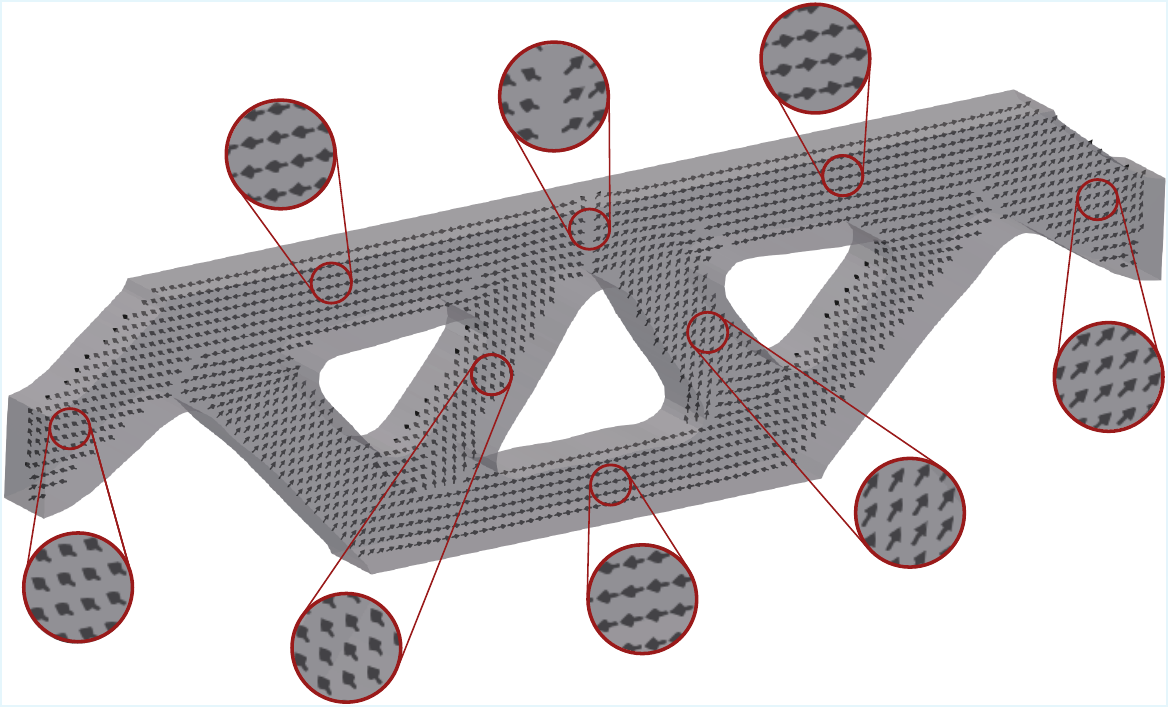}
    \caption{Optimized topology and fiber-orientation for the simply supported beam when both density and fiber directions are considered as design variables. }
    \label{fig:rhotheta}
\end{figure}

\begin{figure}[H]

    \centering
    \includegraphics[width=0.75\linewidth]{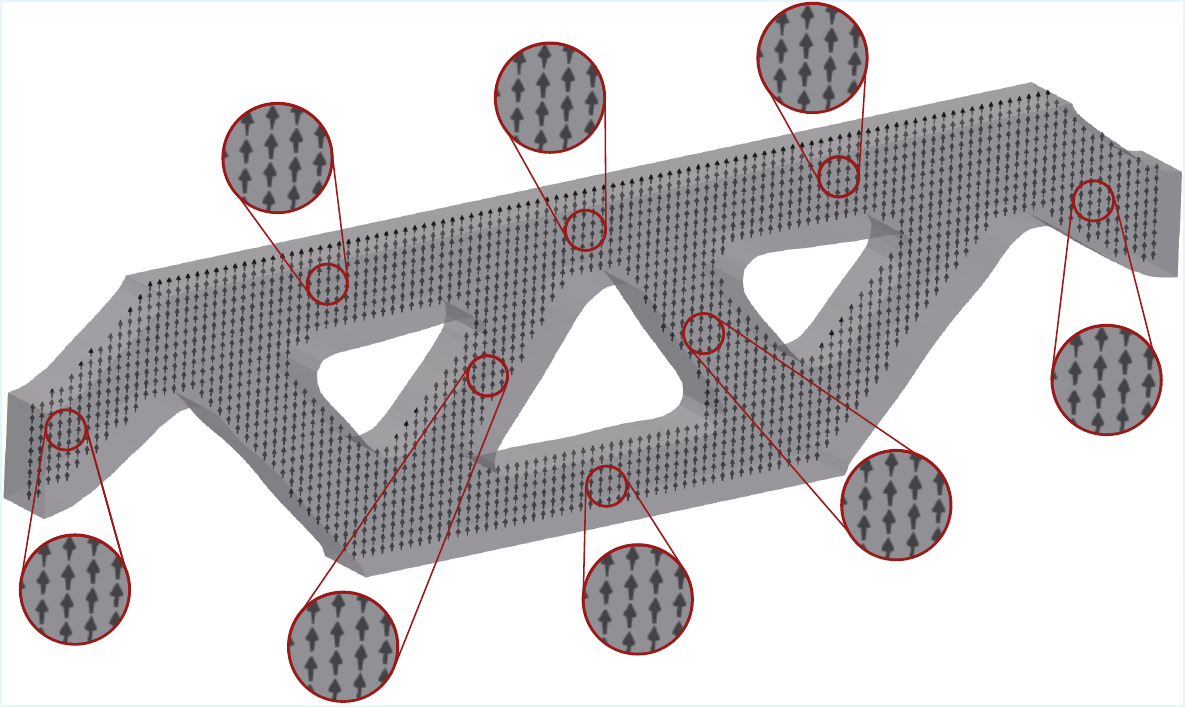}
    \caption{Optimized topology for the simply supported beam when only density is optimized, and fiber directions remain fixed.}
    \label{fig:fixedtheta}
\end{figure}

In this appendix, we show that optimizing the fiber directions along with the topology can have significant differences in the compliance of the optimized structure. We demonstrate on the same problem as in Section ~\ref{sec:fiberRVE} with the simply supported beam. 

To isolate the contribution of fiber-orientation design within the concurrent optimization framework, we repeat the simply supported beam example under two settings:
\begin{itemize}
    \item Topology-only optimization with fixed fiber orientations
    \item Concurrent optimization of topology and fiber orientation
\end{itemize}

In both cases, we assume that the material is anisotropic everywhere, i.e., $\phi = 0.5$ in all elements.

\begin{figure}[H]
    \centering
    \includegraphics[width=0.80\linewidth]{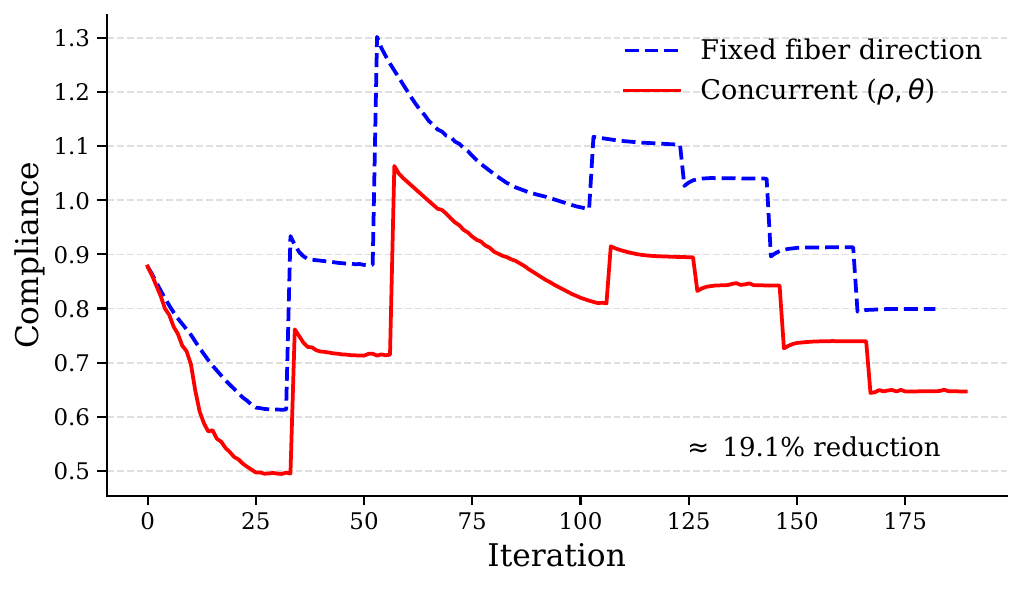}
    \caption{Comparison of compliance evolution between optimization with fixed $\theta$ and concurrent optimization of $\rho$ and $\theta$}
    \label{fig:compliance_compare_appendix}
\end{figure}

Figures~\ref{fig:rhotheta} and \ref{fig:fixedtheta} show the resulting topologies for both cases. When fiber directions are included as design variables, the fibers are optimized to align with the dominant stress trajectories, which increases load transfer and reduces compliance. 

In contrast, when the fiber orientations are fixed, the optimized structure uses more material while resulting in a 19.1\% higher final compliance, as shown in Fig.~\ref{fig:compliance_compare_appendix}.

\section{Geometrical nonlinearity}
\label{app:cantilever_NL}
\begin{figure}
    \centering
    \includegraphics[width=1.0\linewidth]{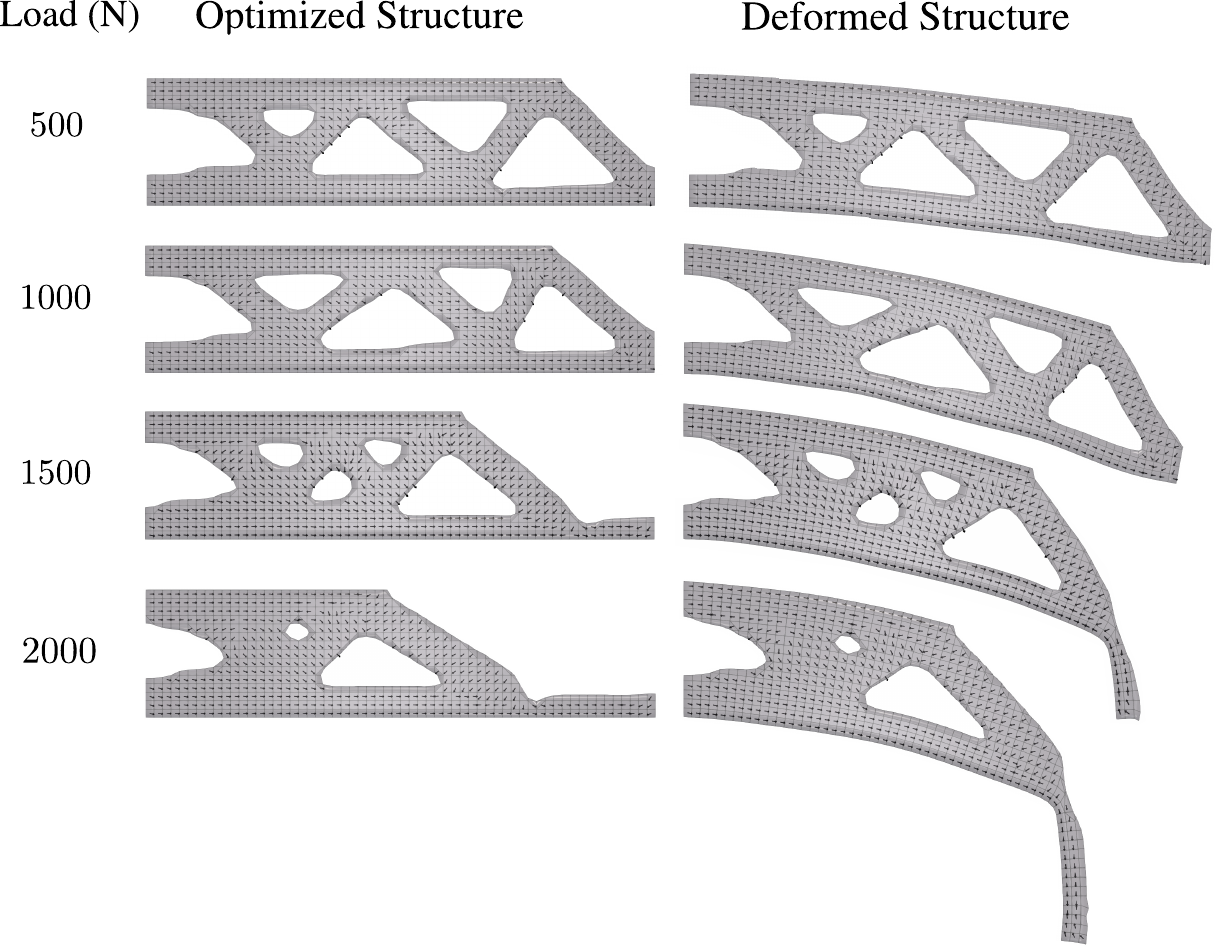}
    \caption{Optimized cantilever beam structures for different load magnitudes, along with their corresponding deformed configurations.}
    \label{fig:cantilever_NL}
\end{figure}

This appendix demonstrates the capability of the proposed PANN-based framework to maintain stability and accuracy in the presence of significant geometrical nonlinearities. We present this by solving the same cantilever beam problem as in Section~\ref{sec:fiberRVE} under increasing load magnitudes. We force the material to be anisotropic everywhere by keeping $\phi=0$ and optimizing material density and fiber orientations. The optimized structures and their corresponding deformed configurations are shown in Fig.~\ref{fig:cantilever_NL}. 

As the applied load increases, the optimized topologies exhibit a clear evolution in their structural layout. For lower load levels, the designs resemble classical linear-elastic cantilever solutions, characterized by relatively symmetric load paths and distributed material layouts. However, as the magnitude of the load increases, geometrical nonlinear effects become significant, leading to pronounced changes in both the optimized topology and the deformations. These observations are consistent with previously reported results in past studies\cite{interpolation_scheme_TO, buhl2000stiffness}, where increasing load levels lead to different optimal designs due to the dominance of geometric nonlinearities. In addition to the evolution of the macroscale topology, the optimized fiber orientations exhibit a clear and physically consistent response to increasing load levels. At lower loads, the fiber directions largely align with the principal stress trajectories of the undeformed configuration, forming continuous reinforcing paths along the dominant load-bearing members. As the load magnitude increases and geometrical nonlinear effects become significant, the fiber orientations progressively adapt to the evolving deformation state. In particular, the fibers remain aligned with the principal stress directions in the deformed configuration, resulting in curved and reoriented patterns that follow the nonlinear load paths. This behavior is especially evident near the fixed end, where bending and rotation are most pronounced, and the fibers form continuous bands that track the changing stress directions.

\clearpage
\bibliography{references}

\end{document}